\documentclass{article}

\usepackage{arxiv}

\usepackage[utf8]{inputenc} % allow utf-8 input
\usepackage[T1]{fontenc}    % use 8-bit T1 fonts
\usepackage{hyperref}       % hyperlinks
\usepackage{url}            % simple URL typesetting
\usepackage{booktabs}       % professional-quality tables
\usepackage{amsfonts}       % blackboard math symbols
\usepackage{nicefrac}       % compact symbols for 1/2, etc.
\usepackage{microtype}      % microtypography
\usepackage{lipsum}		% Can be removed after putting your text content
\usepackage{graphicx}
\usepackage{natbib}
\usepackage{doi}
\usepackage{caption}
\usepackage{subcaption}
\usepackage{algorithm}
\usepackage{algorithmic}

\title{An Interactive Web-Based System for Creating Single Panel Cartoons with Visually Valid Compositions}

\author{ \href{https://orcid.org/0000-0003-3618-4166}{\includegraphics[scale=0.06]{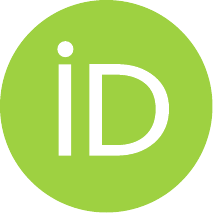}\hspace{1mm}Ergun Akleman}\thanks{Joint with Computer Science and Engineering Department.} \\
	Visual Computing \& Computational Media,\\ Texas A\&M University, College Station, TX, 77831\\
	\texttt{ergun@tamu.edu} \\
	\And
	Akhilesh Vijaykumar\\
	Visualization Department, \\ Texas A\&M University, College Station, TX, 77831\\
	\texttt{akhivijay97@tamu.edu } \\
     \And
	Richard Furuta \\
	Computer Science and Engineering Department, \\ Texas A\&M University, College Station, TX, 77831\\
	\texttt{furuta@tamu.edu} \\
     \And
	Derya Akleman \\
	Statistics Department, \\ Texas A\&M University, College Station, TX, 77831\\
	\texttt{akleman@tamu.edu} \\
}

\renewcommand{\shorttitle}{Visual Valid Single Panel Cartoon Compositions}

\hypersetup{
pdftitle={A template for the arxiv style},
pdfsubject={q-bio.NC, q-bio.QM},
pdfauthor={David S.~Hippocampus, Elias D.~Striatum},
pdfkeywords={First keyword, Second keyword, More},
}

\begin{document}
\maketitle
	
\begin{abstract}

The creation of cartoon-based stories (comics) requires a lot of creativity and hard work for naive users \cite{akleman2015theoretical}. We observe that single-panel cartoons are the building blocks of any comic story. To develop a strong comic story, it is critical to obtain visually valid single panels \cite{liu2012never}. In this work, we have developed a methodology to guarantee the placement of characters to obtain a valid cartoon frame based on the methods used by professional cartoonists. Using this methodology, we have developed a web-based system to create single-panel cartoons from a given set of character images. We have made this system available in GitHub as open-source so that this basic single-panel cartoon can be used as infrastructure to develop more complex structures. Our web-based system for single-panel cartoons can be viewed at http://storytelling.viz.tamu.edu.

\end{abstract}
	
	
	\begin{figure}
		\begin{tabular}{ccccc}
		\fbox{\includegraphics[width=0.22\linewidth]{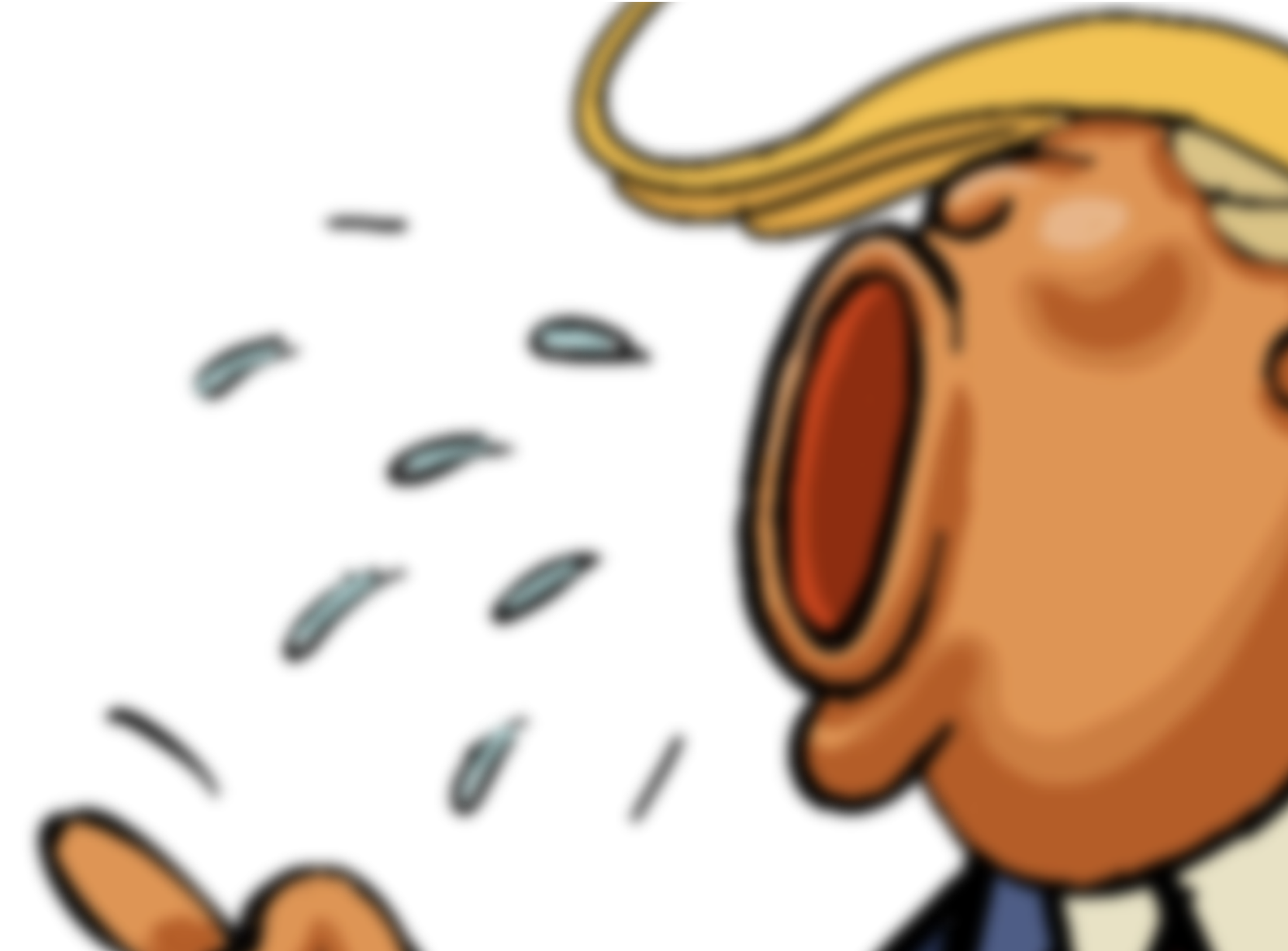}}&
		\fbox{\includegraphics[width=0.22\linewidth]{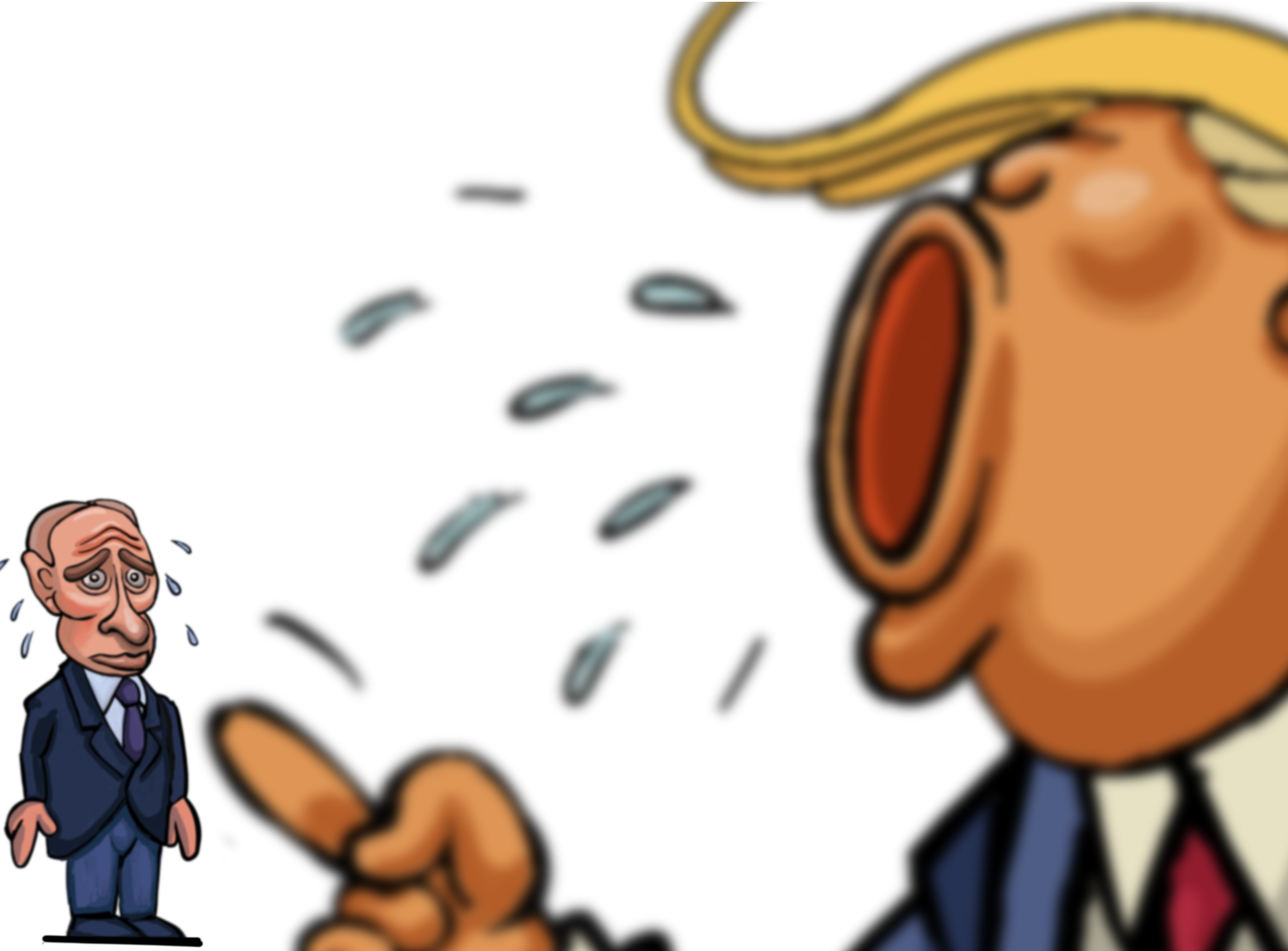}}&
		%\fbox{\includegraphics[width=0.22\linewidth]{Views/x01}}&
		\fbox{\includegraphics[width=0.22\linewidth]{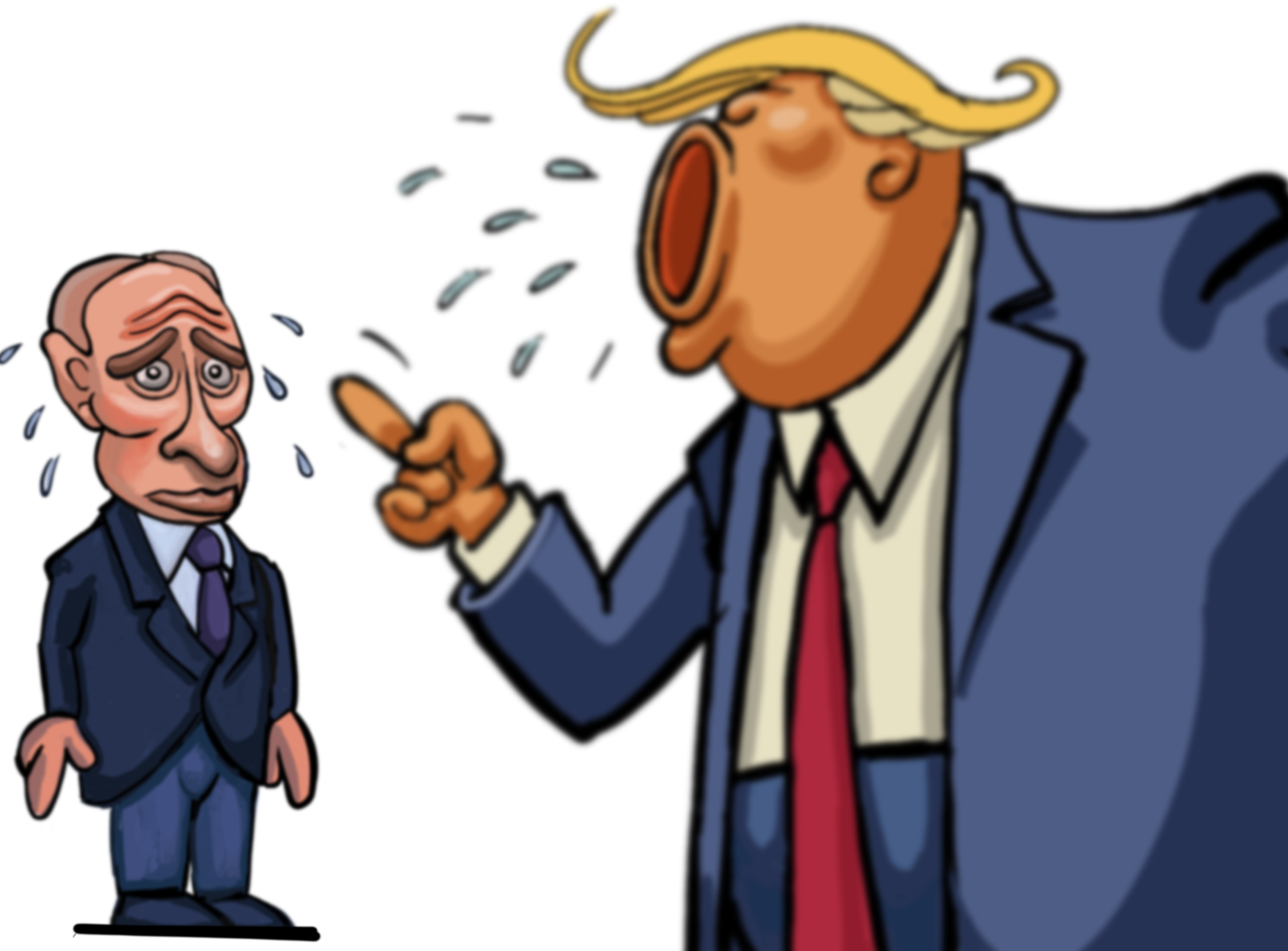}}&
		\fbox{\includegraphics[width=0.22\linewidth]{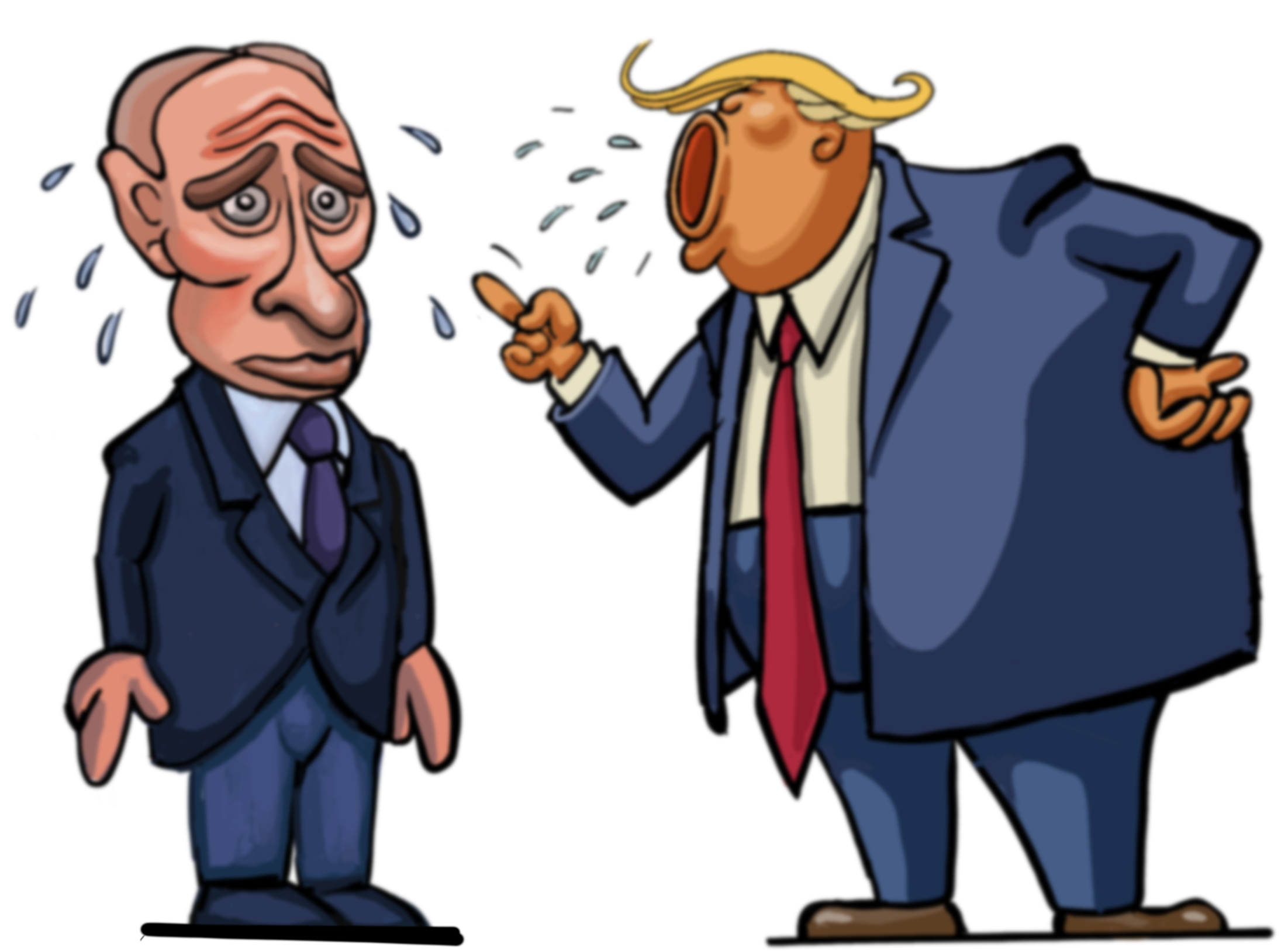}}\\
		(A) Internal Left & 
		(B) External Left 1&
		(C) External Left 2 &
		(D) Apex \\
		\fbox{\includegraphics[width=0.22\linewidth]{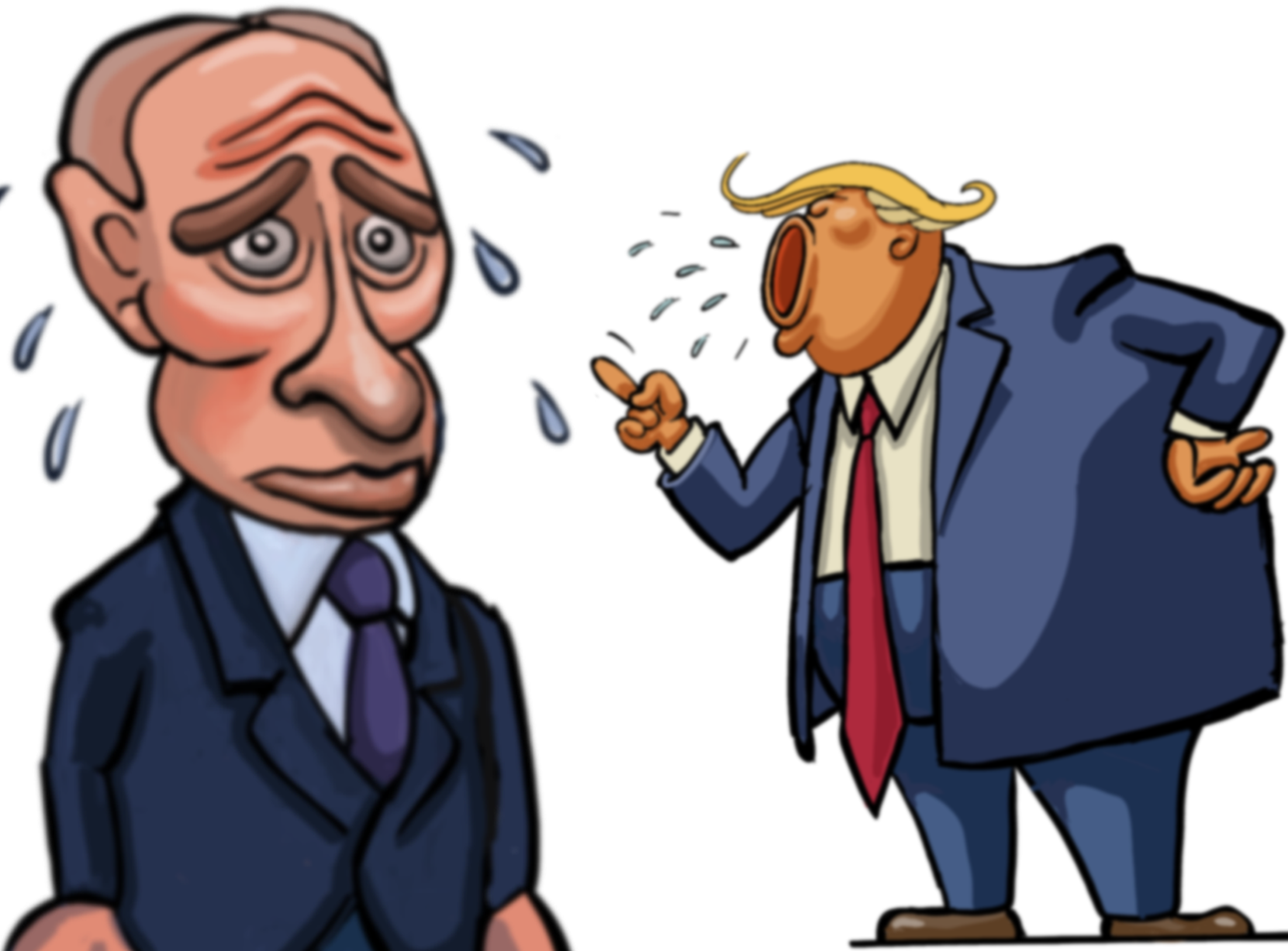}}&
		\fbox{\includegraphics[width=0.22\linewidth]{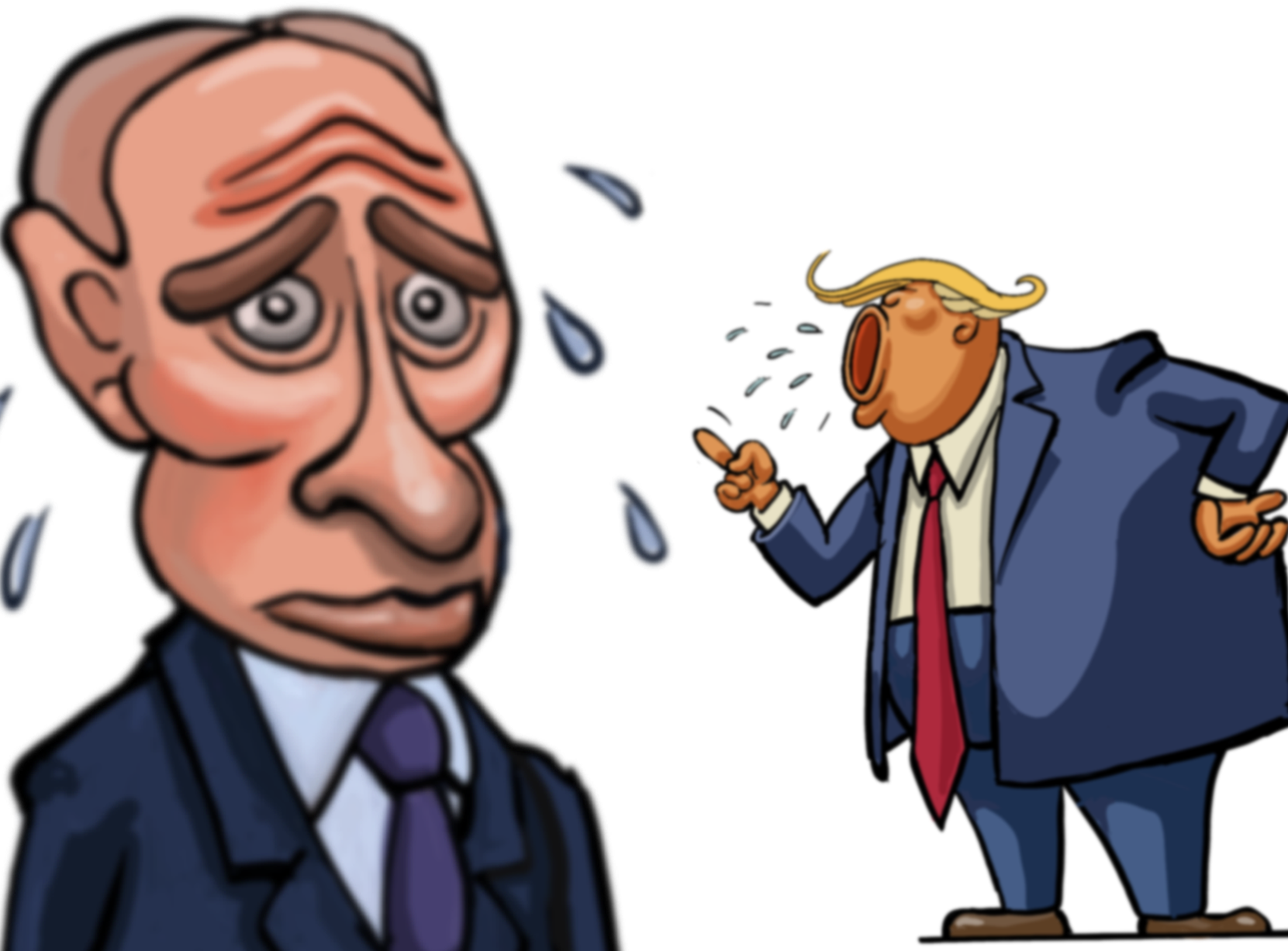}}&
		%\fbox{\includegraphics[width=0.22\linewidth]{Views/x13}}&
		\fbox{\includegraphics[width=0.22\linewidth]{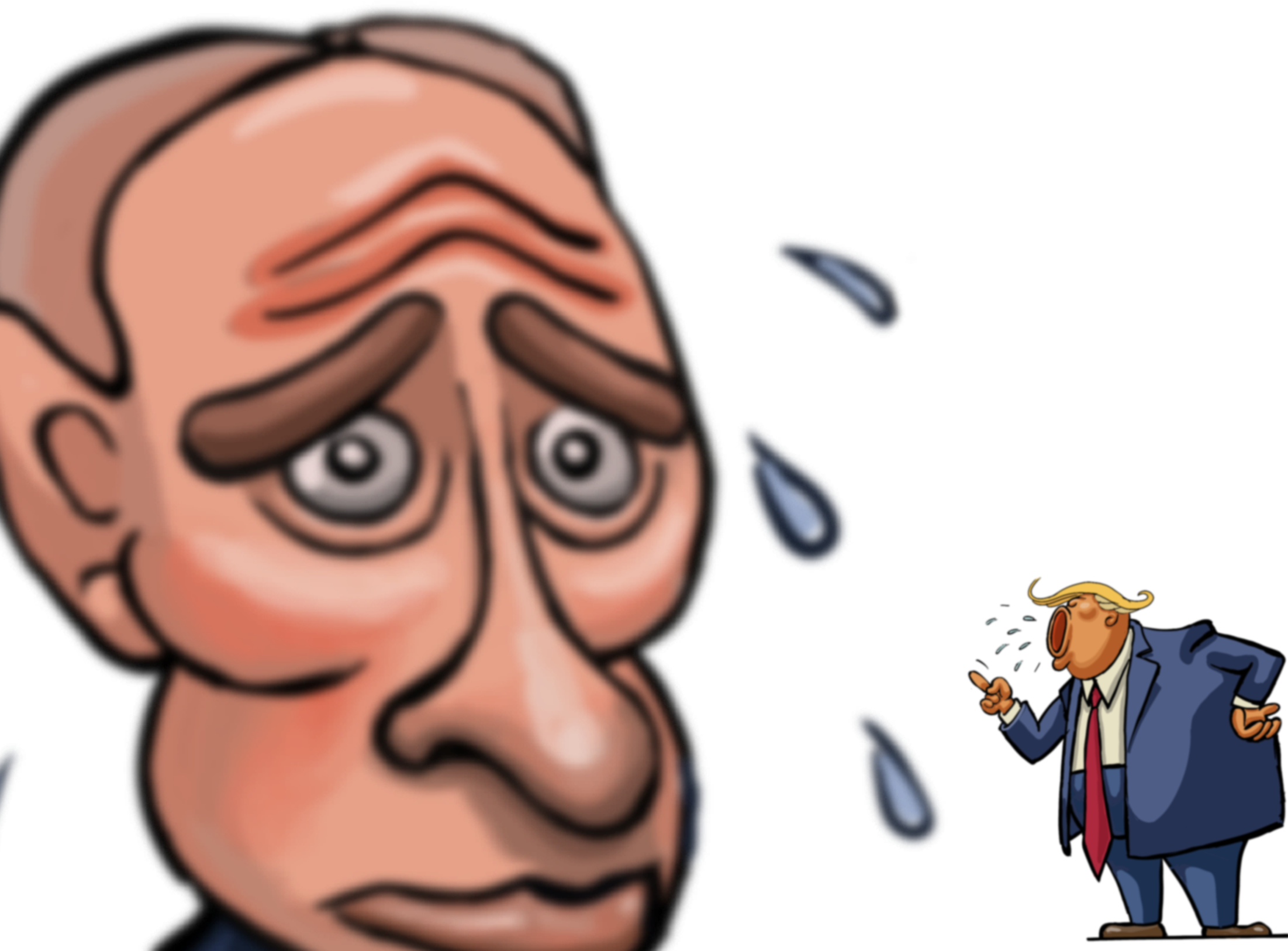}}&
		\fbox{\includegraphics[width=0.22\linewidth]{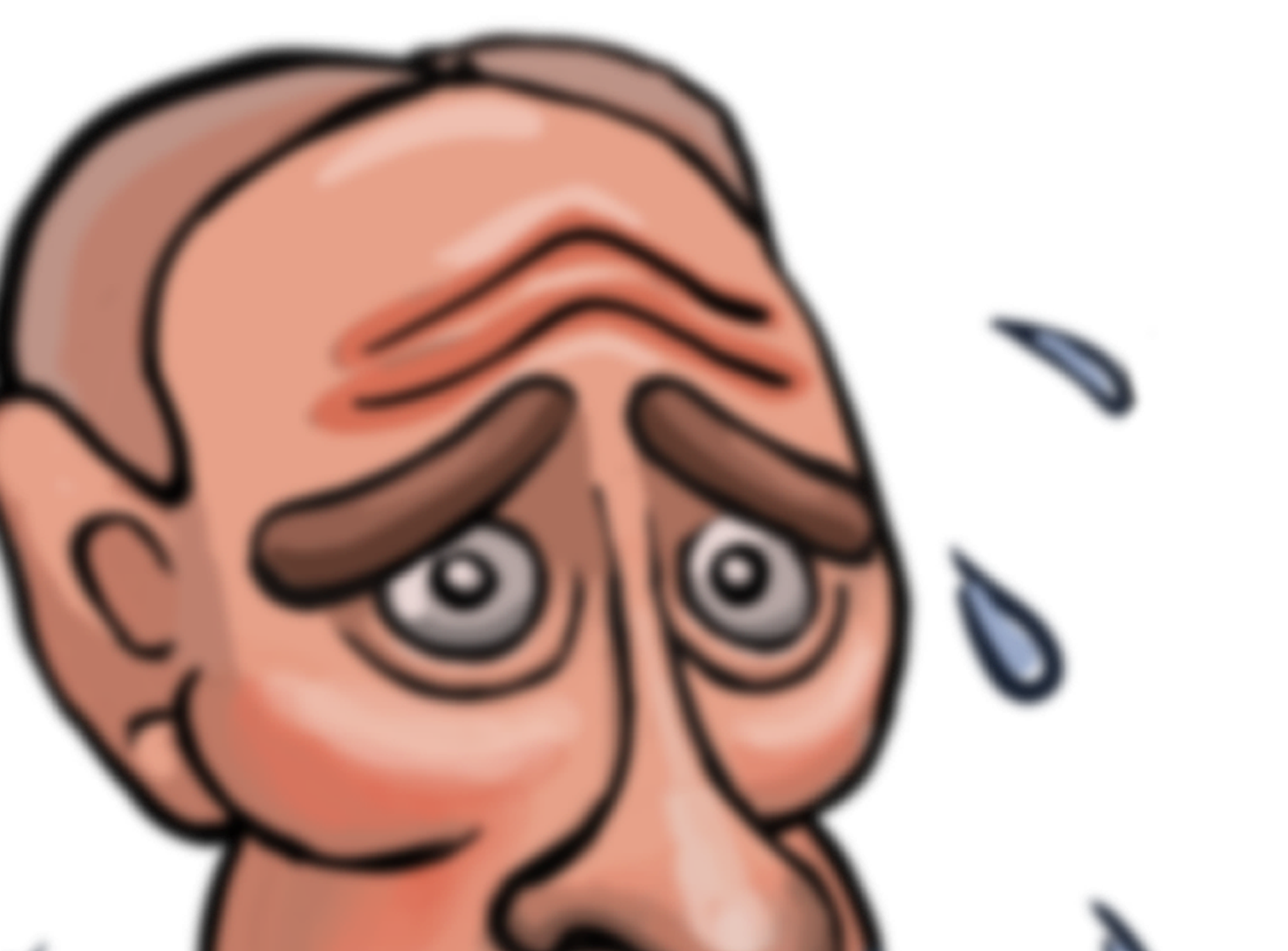}}\\
		(F) External Right 1& 
		(G) External Right 2&
		(H) External Right 3 &
		(I) Internal Right\\
		\end{tabular}
		\caption{\it An example of character placement in a single panel with two characters using our single parameter solution. This one-parameter character placement is based on a continuous interpolation of conceptual 2D versions of 3D camera views: from internal view to external view and from external view to apex. }
		\label{fig:teaser}
	\end{figure}

\begin{figure*}[htbp!]
\centering		\fbox{\includegraphics[width=0.60\linewidth]{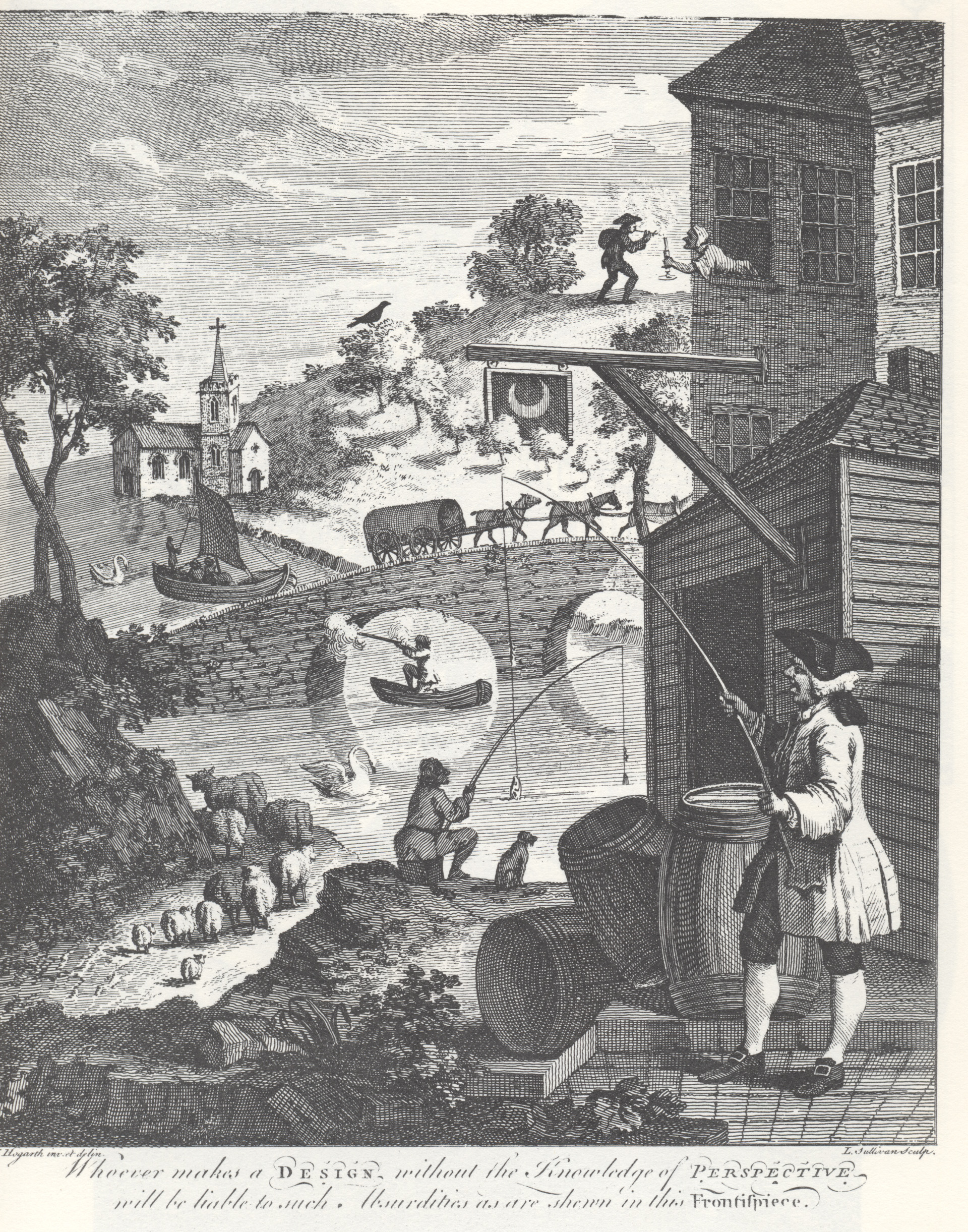}}
		\caption{\it Our motivation for visual validity came from William Hogart's Engraving called \textit{``Satire on False Perspective}. Hogart demonstrated that anybody who does not have expert illustration knowledge is liable to create non-valid compositions.  }
		\label{fig:hogart}	
\end{figure*}

\section{Introduction and Motivation}

Comics and cartoons as a distinct form of art emerged during $17^{th}$ and $18^{th}$ centuries by great artists such as William Hogarth, James Gillray, Thomas Rowlandson, and George Cruikshank \cite{smolderen2014origins,rowson2015satire,horn1976world}. Despite their long history, comics and cartoons are one of the least understood art forms that require not only a significant amount of illustration and drawing experience but also a good understanding of visual storytelling. 
One of the key elements of cartoons and comics is the creation of valid single cartoon or comic panels that can provide strong visual storytelling experience
\cite{carrier2001aesthetics,eisner2008comics,mccloud1993understanding,mcloud2000reinventing,mccloud2006making,akleman2020}. There exist cartoon interfaces such as ToonyTool \cite{schoonhoven2021} that provide complete freedom to place images in a panel. Users translate, scale, and rotate multiple images to produce a single panel. However, with such complete freedom, it is hard for naive users to compose visually valid panels that can be visually compelling.  There is \textit{a need} for a simple approach to guarantee that such naive users can obtain visually valid and compelling panels by specifying only a few parameters. 

The concept of visual validity was introduced by English painter and cartoonist William Hogarth (1697-1764) in his 1754 engraving called \textit{``Satire on False Perspective''} \cite{cornew2001perspectival,hogart1754}. In the subtitle, Hogarth wrote that \textit{``Whoever makes a Design without the Knowledge of Perspective will be liable to such Absurdities as are shown in this [engraving]''}. There are more than 20 \textit{Absurdities} in this image and it is even hard to find all absurdities. In this paper, we call these absurdities visually non-valid compositions. Our goal is to guarantee the creation of the only visually valid compositions. 

In this work, our major contribution is the identification of a single parameter family of visually valid single-panel cartoon compositions based on professional cartoon expertise. Using this family, we have identified an approach to obtain appealing single cartoon panels using only one parameter and developed a web-based prototype. Our approach is based on five essential 3D camera views. These five views are widely used in movies to create dynamic dialogues. We observe that these particular camera views can be adopted for 2D character placement. Based on this observation, we have identified 2D conceptual versions of these five views for any given two images that represent characters/objects. Our one-parameter solution is a continuous interpolation between these five essential views. Our solution can be obtained simply by translation and scaling of these two images. The operation can recursively be extended to any number of images by creating a binary tree.

\begin{figure*}[htbp!]
	\begin{tabular}{cccccccccc}
		\fbox{\includegraphics[width=0.32\linewidth]{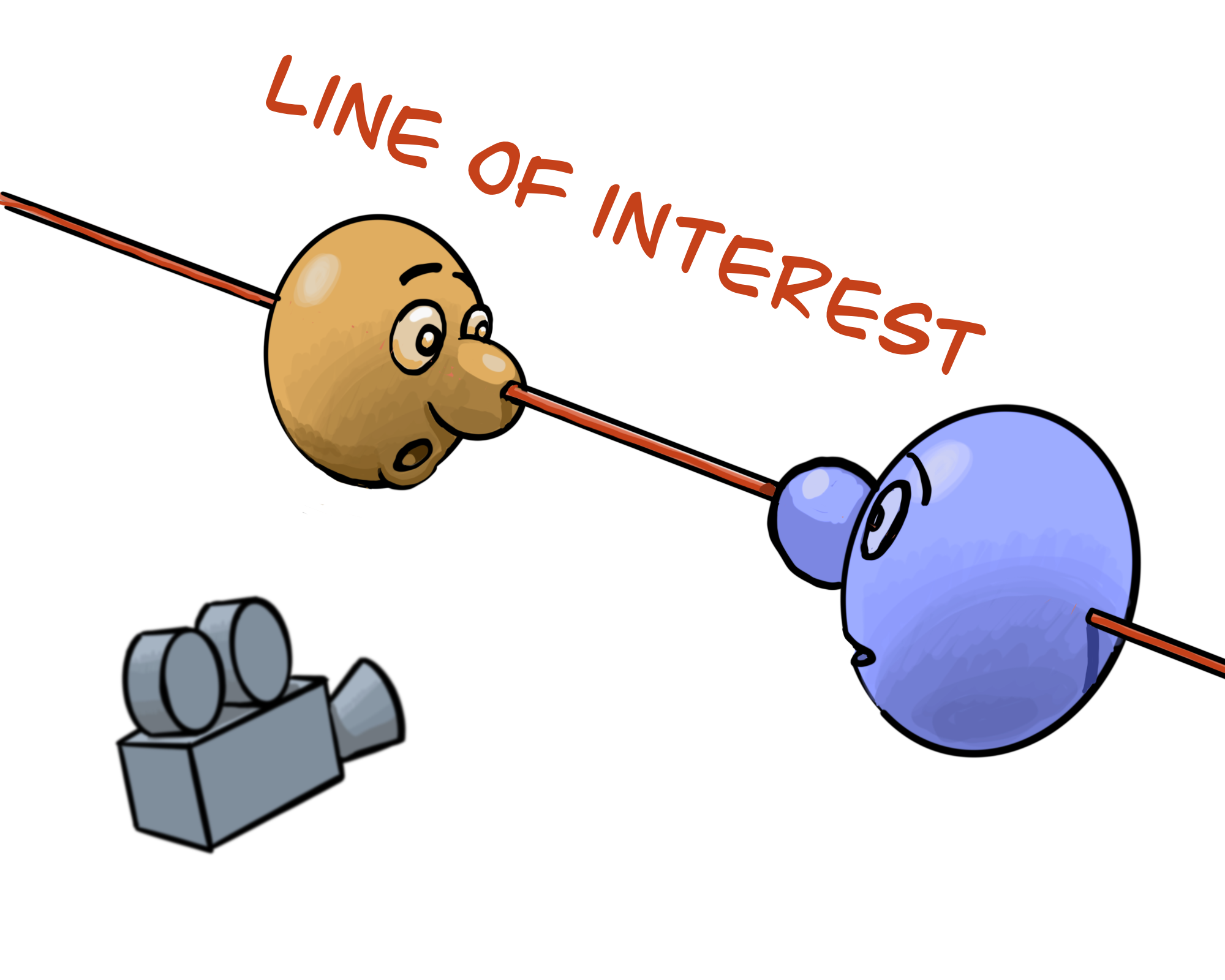}}&
		\hfill
       \fbox{ \includegraphics[width=0.26\linewidth]{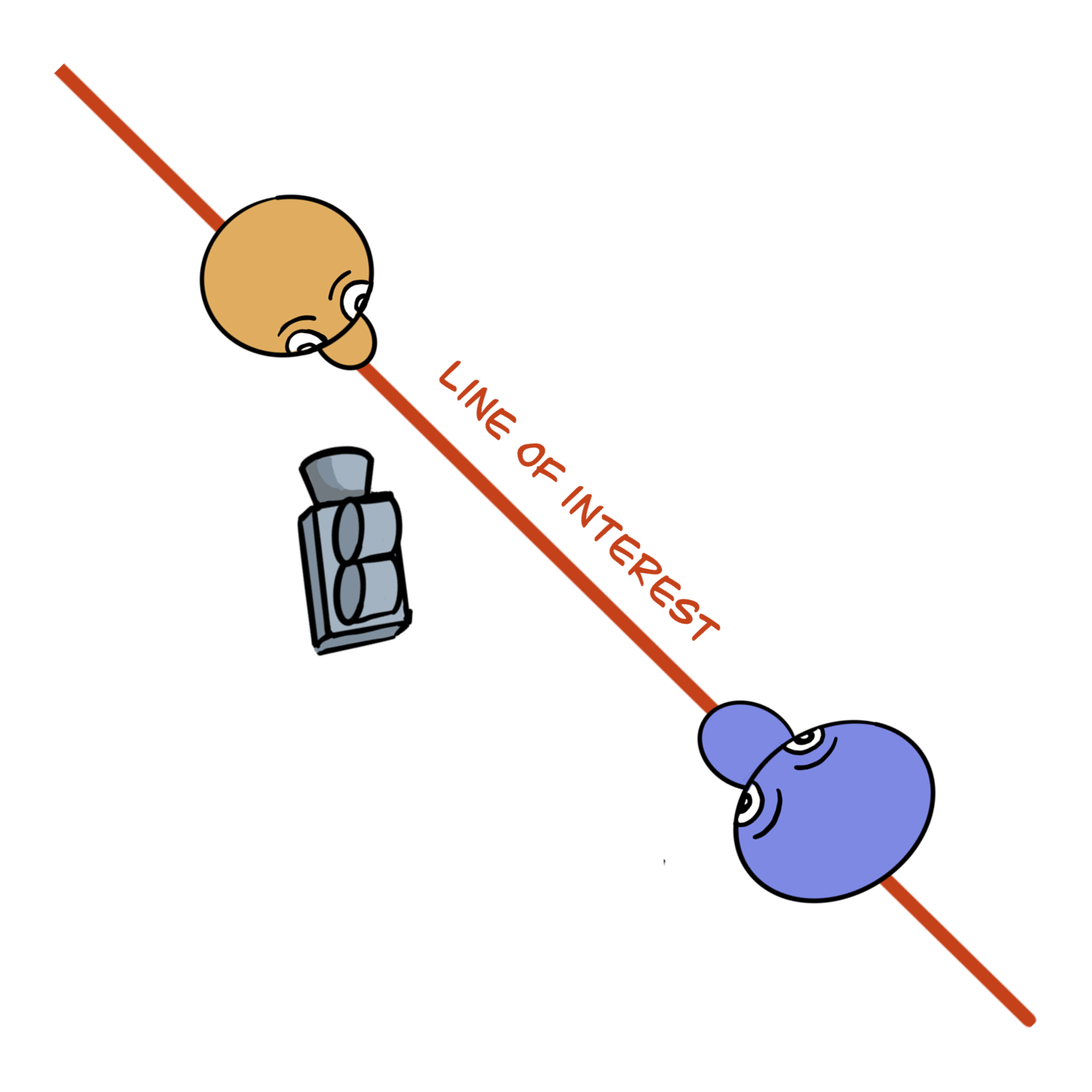}}&
	 \hfill
	 	\fbox{	\includegraphics[width=0.26\linewidth]{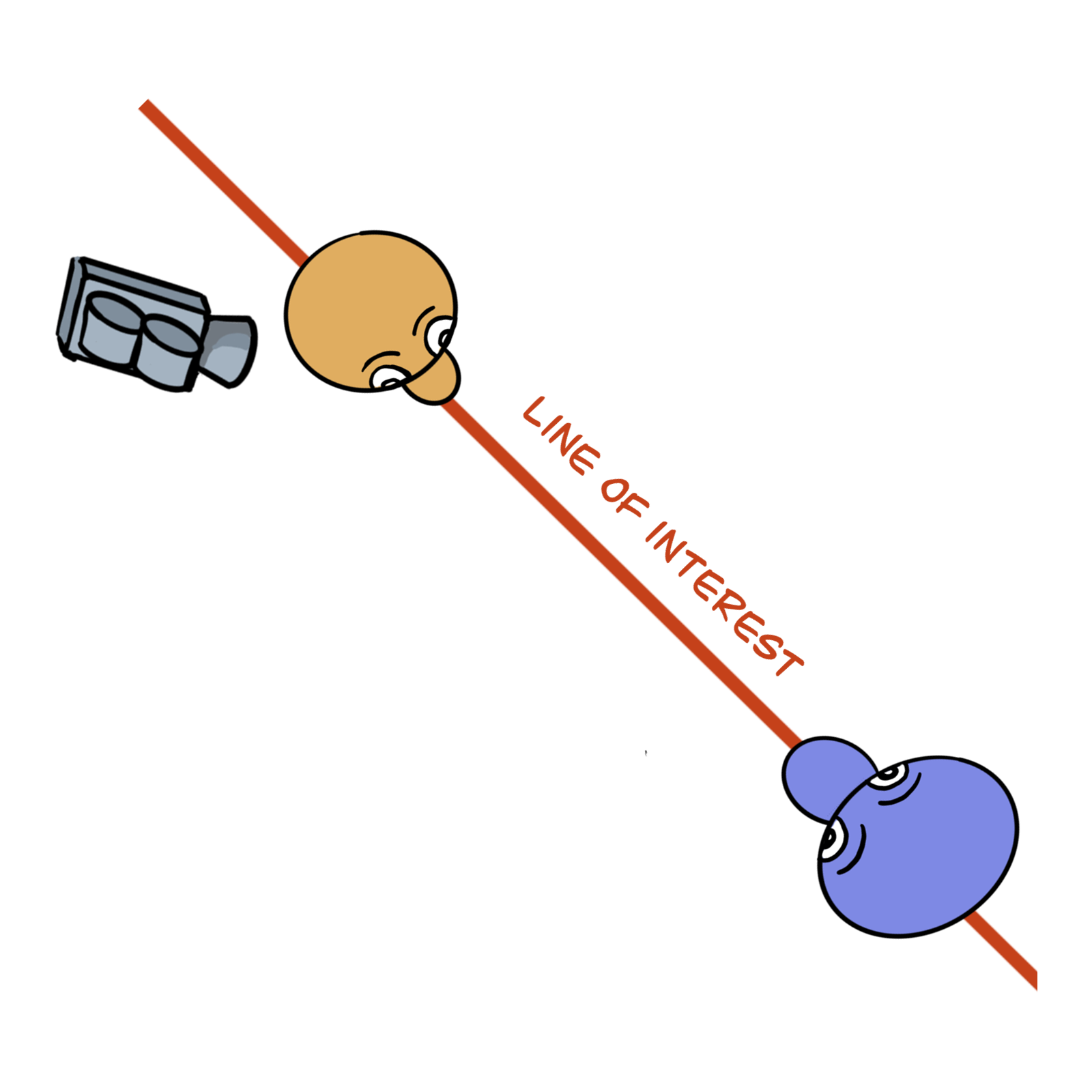}}&
\\
    (A) Line of Interest. &
      (B)  Internal Left.&
    (C) External Left. \\
		\fbox{\includegraphics[width=0.26\linewidth]{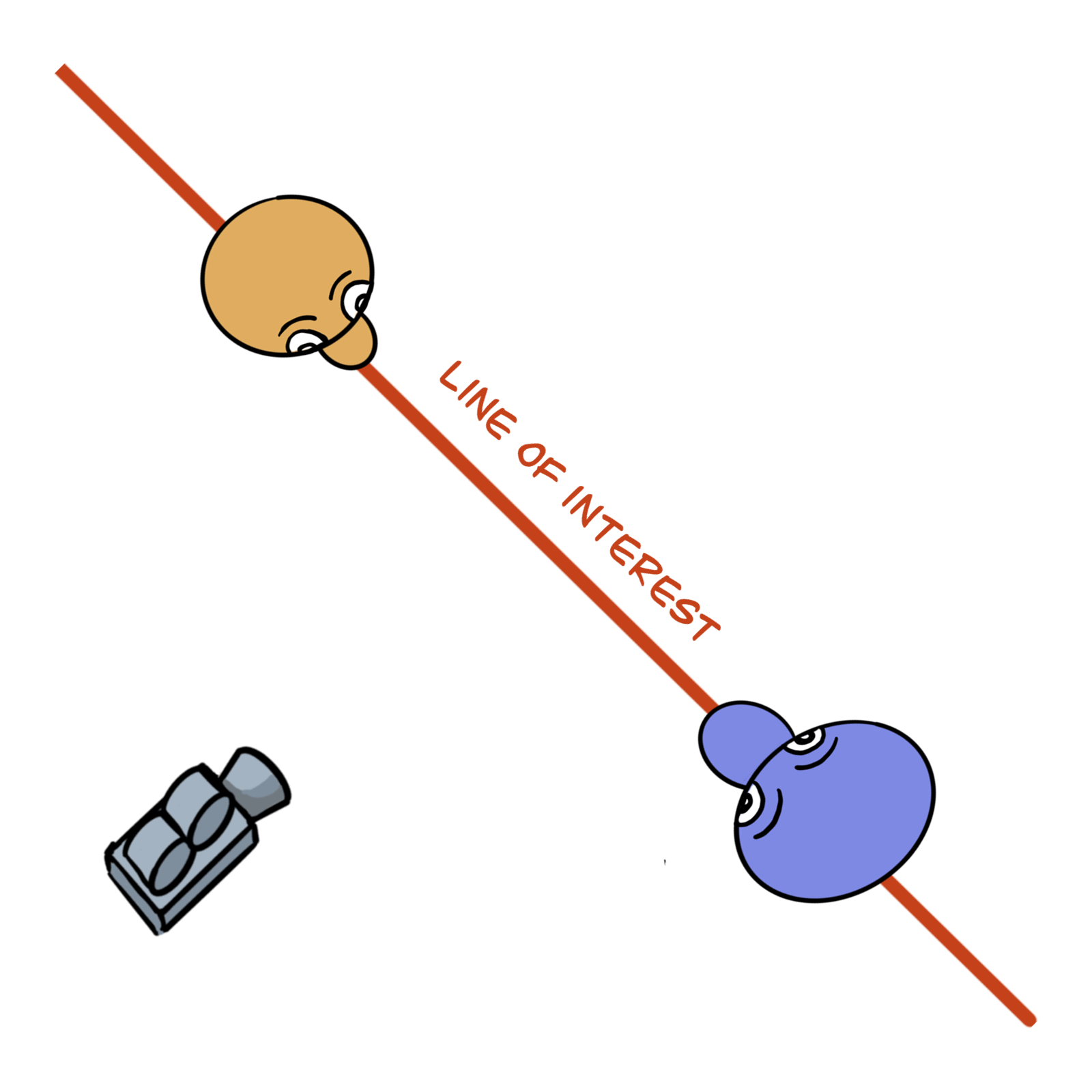}}&
    \hfill
			\fbox{\includegraphics[width=0.26\linewidth]{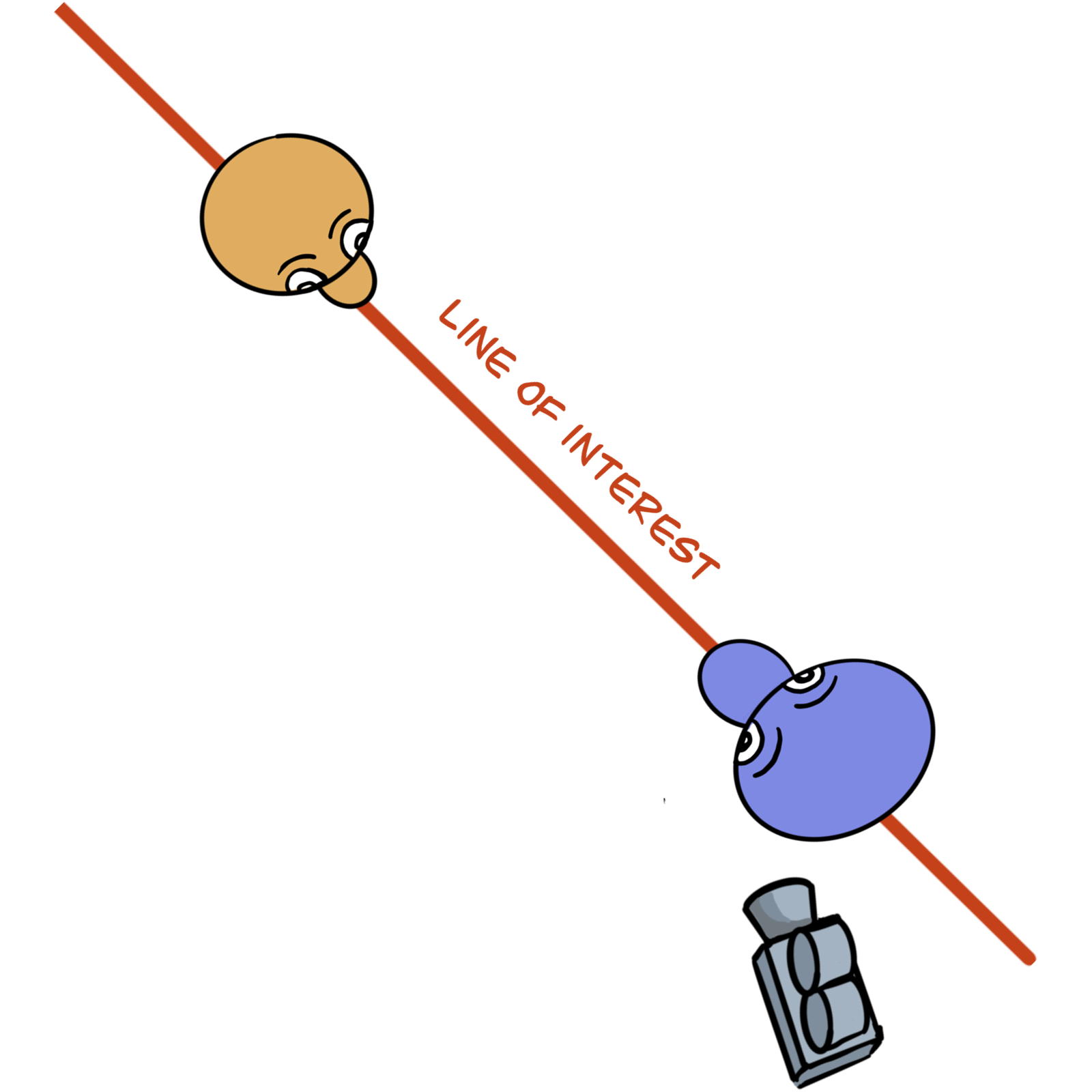}}&
    \hfill	
    \fbox{\includegraphics[width=0.26\linewidth]{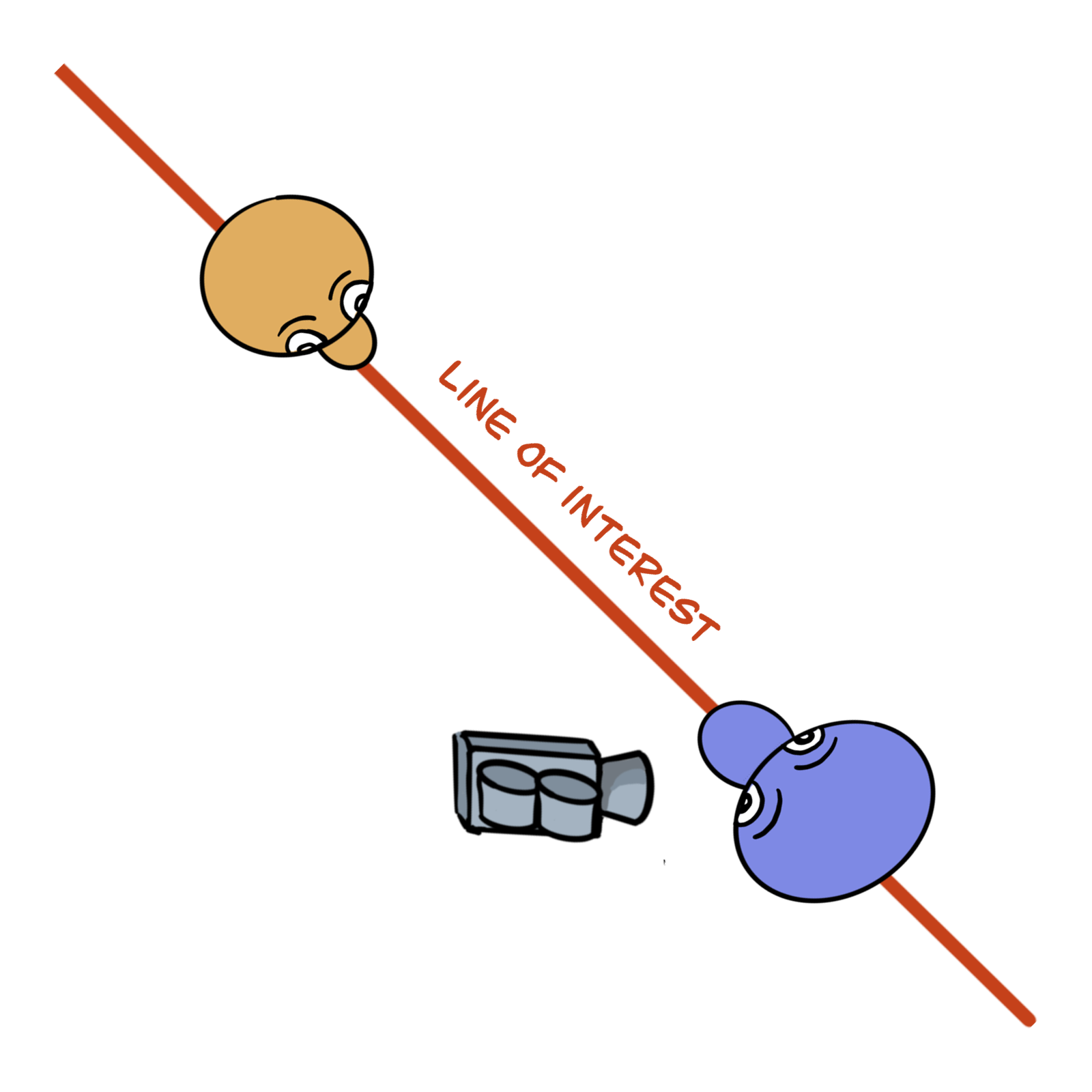}}\\
    (E) Apex.  &
    (F) External Right. &
     (G) Internal Right.  \\
    \end{tabular}
		\caption{\it In 3D, any two objects/characters define a a line of interest as shown on A. If the camera stays only one side of the line of interest, the objects/characters consistently stay at the left and right of the panel. In this case, the yellow object/character will always stay at the left and the blue object/character will always stay at the right. There are only five qualitatively distinct camera positions as shown from (B) to (G). We have observed that we can create a continuous path between 2D conceptual views that corresponds to these five camera views. }
		\label{fig:CameraView}	
\end{figure*}

\section{On the Importance of 3D Camera Views For Storytelling}

Five qualitatively different 3D views, internal left view, external left view, apex view, external right view, and internal right view, are widely used in movies and comics to create dynamic dialogues \cite{arijon1991grammar,he1996virtual} (see Figure~\ref{fig:CameraView}).
We observed that these five views are key elements to analyzing any movie and photographic scene in terms of the storytelling point of view and producing visually valid compositions in cartoons. Assume that the scene has two essential elements, objects or characters, that are given by their positions $\mathbf{p}_1$ and $\mathbf{p}_2$; or a motion that is given a position and direction vector, $\mathbf{p}_1$ and $\vec{v}_2$. These two can be considered exactly the same way by assuming  $\vec{v}_2=\mathbf{p}_2-\mathbf{p}_1$. Now, let the camera position be given by $\mathbf{p}_0$.

\begin{figure}[htbp!]
	\begin{tabular}{cccccccccc}
		\fbox{\includegraphics[width=0.49\linewidth]{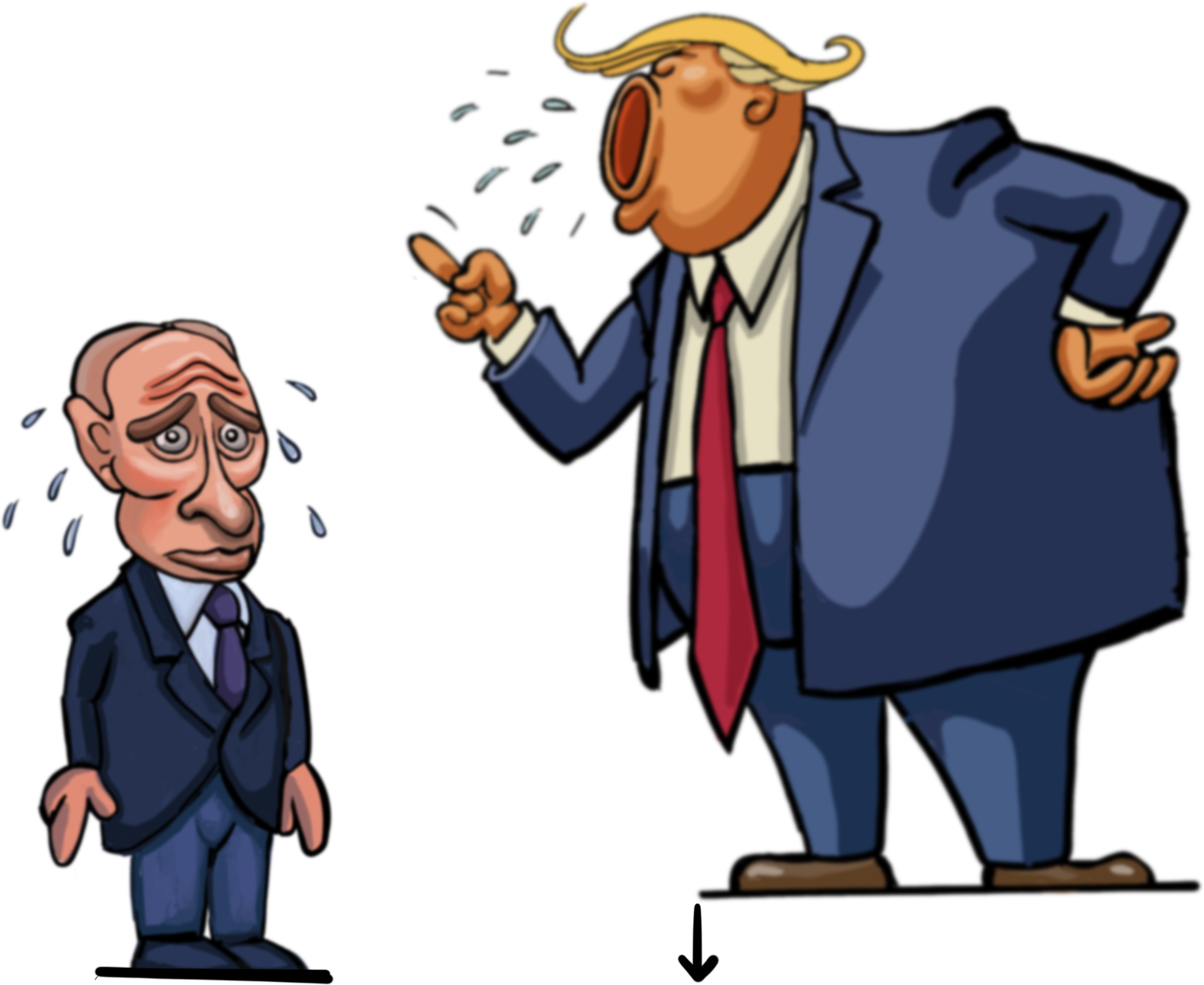}}&
		\hfill
       \fbox{ \includegraphics[width=0.42\linewidth]{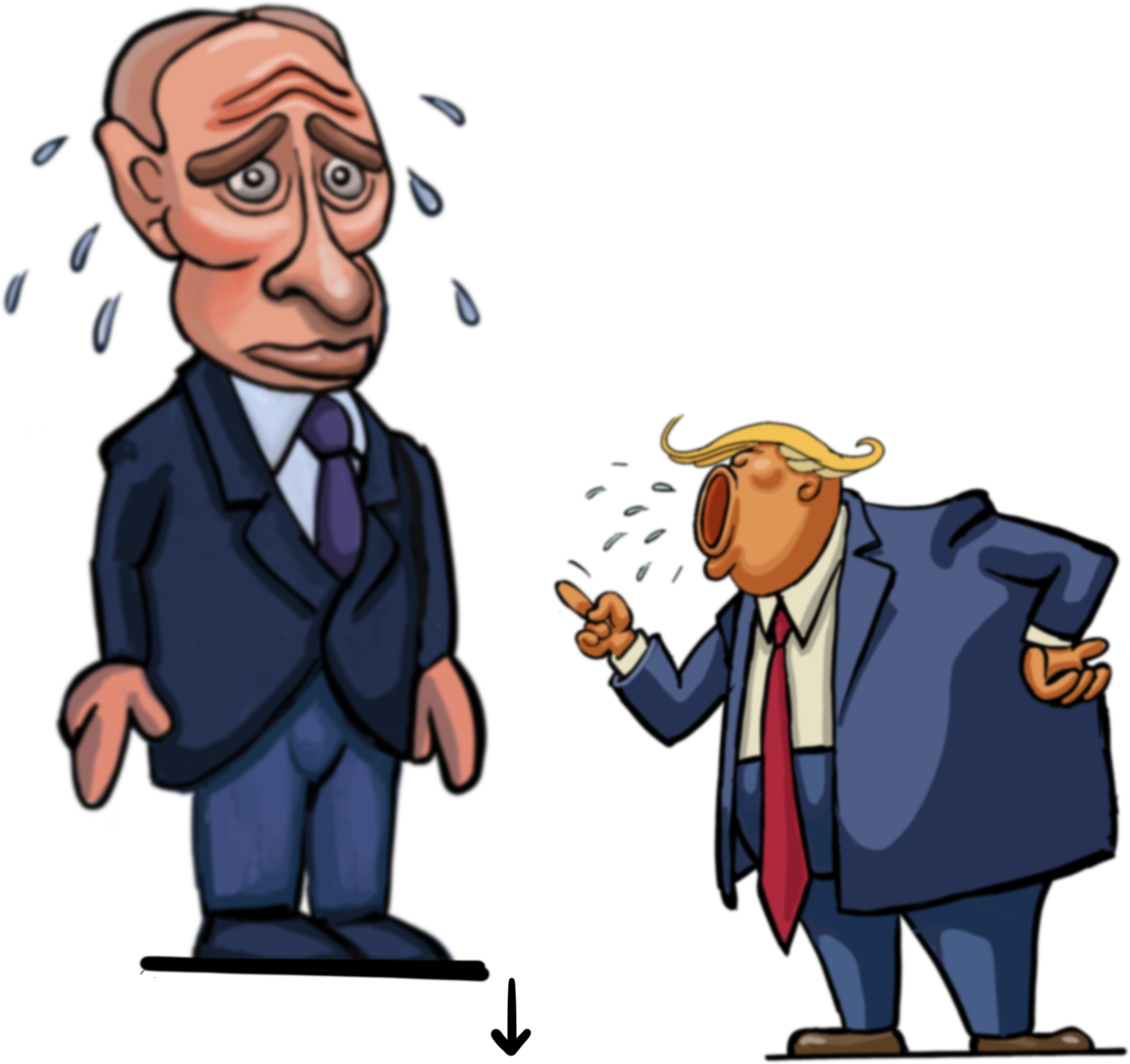}}\\
    \end{tabular}
		\caption{\it Two examples of visually non-valid compositions. }
		\label{fig:Unacceptable}	
\end{figure}

If $|\mathbf{p}_1-\mathbf{p}_0|=|\mathbf{p}_2-\mathbf{p}_0|$, we call this an apex view. In this case, the two objects/characters have exactly the same importance. On the other hand, If $|\mathbf{p}_1-\mathbf{p}_0|>|\mathbf{p}_2-\mathbf{p}_0|$, the camera is closer to the object/character 1 and object/character 1 is emphasized. However, there are two possible cases based on the orientation of the camera: (1) External view, in which object 2 is still visible but smaller; and (2) Internal view, in which object 2 is outside of the frame and is invisible in the frame. In other words, although the external view emphasizes one of the object(s)/character(s) by giving it more screen space, the internal view has even stronger emphasis since that particular object/character is the center of attention by being completely alone in the scene. 

Based on these five views, we can control the emphasis given any object or character; even a single frame. Assume that we have a human and a key. We can make either the human more important or the key more important in a single frame. When we have multiple frames, different views create an additional dynamic and the order of images creates a story that can easily be interpreted by the viewer. Note that the analysis can be applied to multiple objects and characters by grouping them using a binary tree. For 2D scene generation, we must also apply the same rules as if there exists a conceptual camera that can help us decipher the meaning of the image.

\section{2D Conceptual Versions of 3D Camera Views}

Let $I_{i,j}$ denote $j^{th}$ image of $i^{th}$ object/character given with transparent background. We also assume that we only know the width and height of each of these images, denoted by $w_{i,j}$ and $h_{i,j}$ respectively. Each image is expected to be tightly packed in the sense that there is no space left at the top, bottom, left, and right. We also expect that objects or characters can only be cut from the bottom. Our goal is to place them in a reasonably successful way with this limited amount of rules to produce visually valid compositions. 
\begin{figure}[htbp!]
	\begin{tabular}{cccccccccc}
		\fbox{\includegraphics[width=0.49\linewidth]{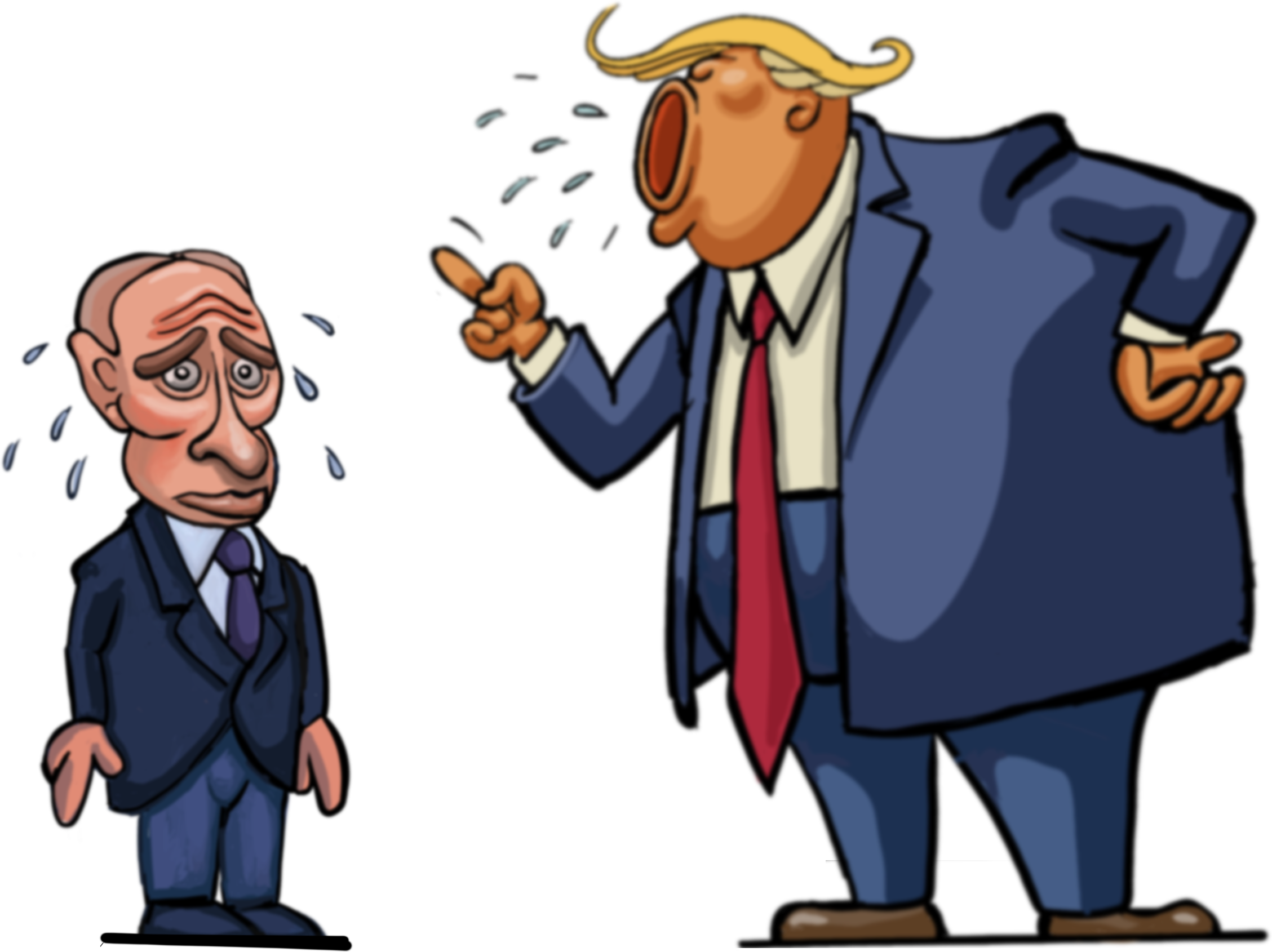}}&
		\hfill
       \fbox{ \includegraphics[width=0.42\linewidth]{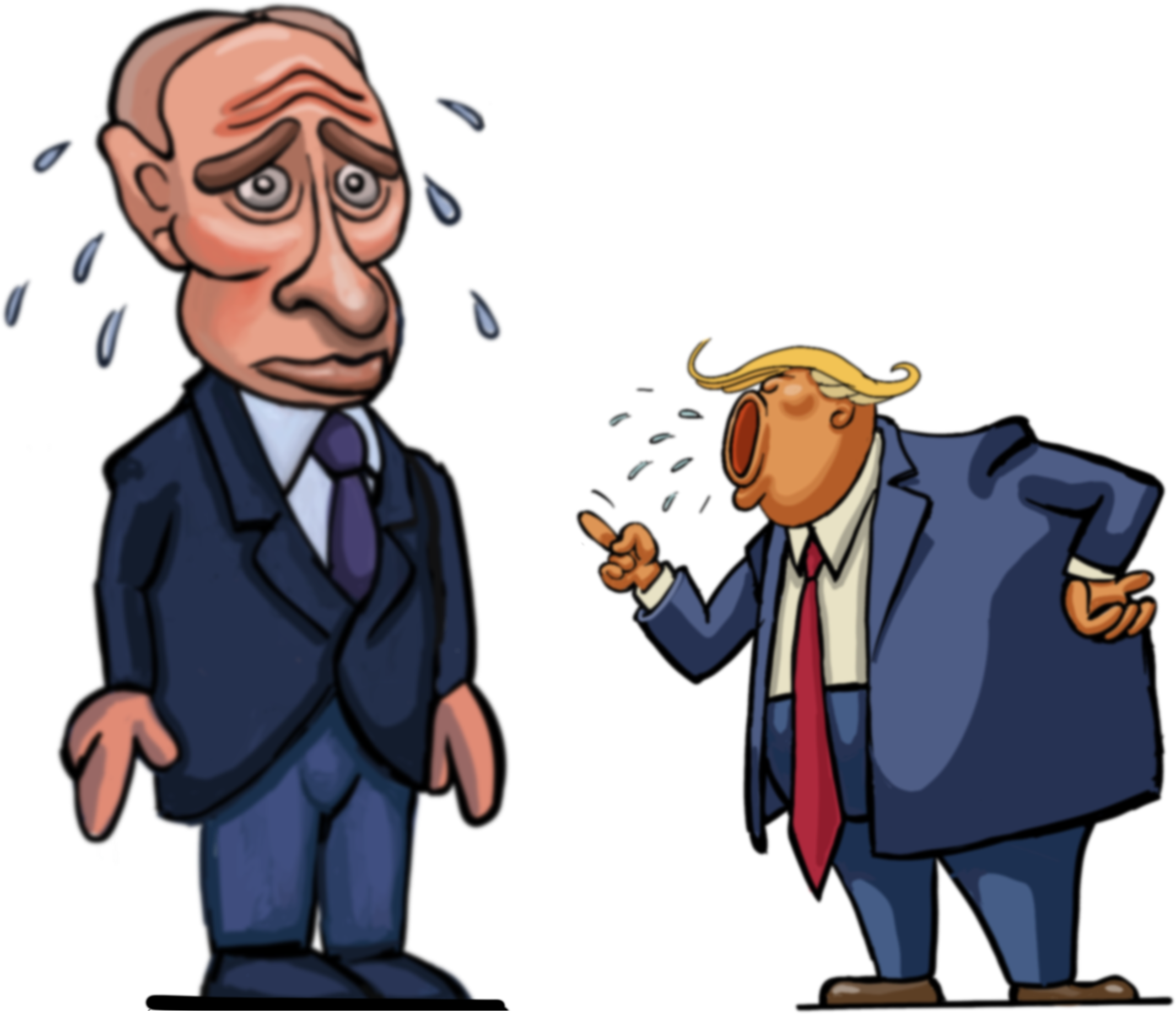}}\\
    \end{tabular}
		\caption{\it Two examples of visually valid compositions that can be used as Apex views. }
		\label{fig:Acceptable}	
\end{figure}

\begin{figure}[htbp!]
	\begin{tabular}{cccc}
       \fbox{ \includegraphics[width=0.39\linewidth]{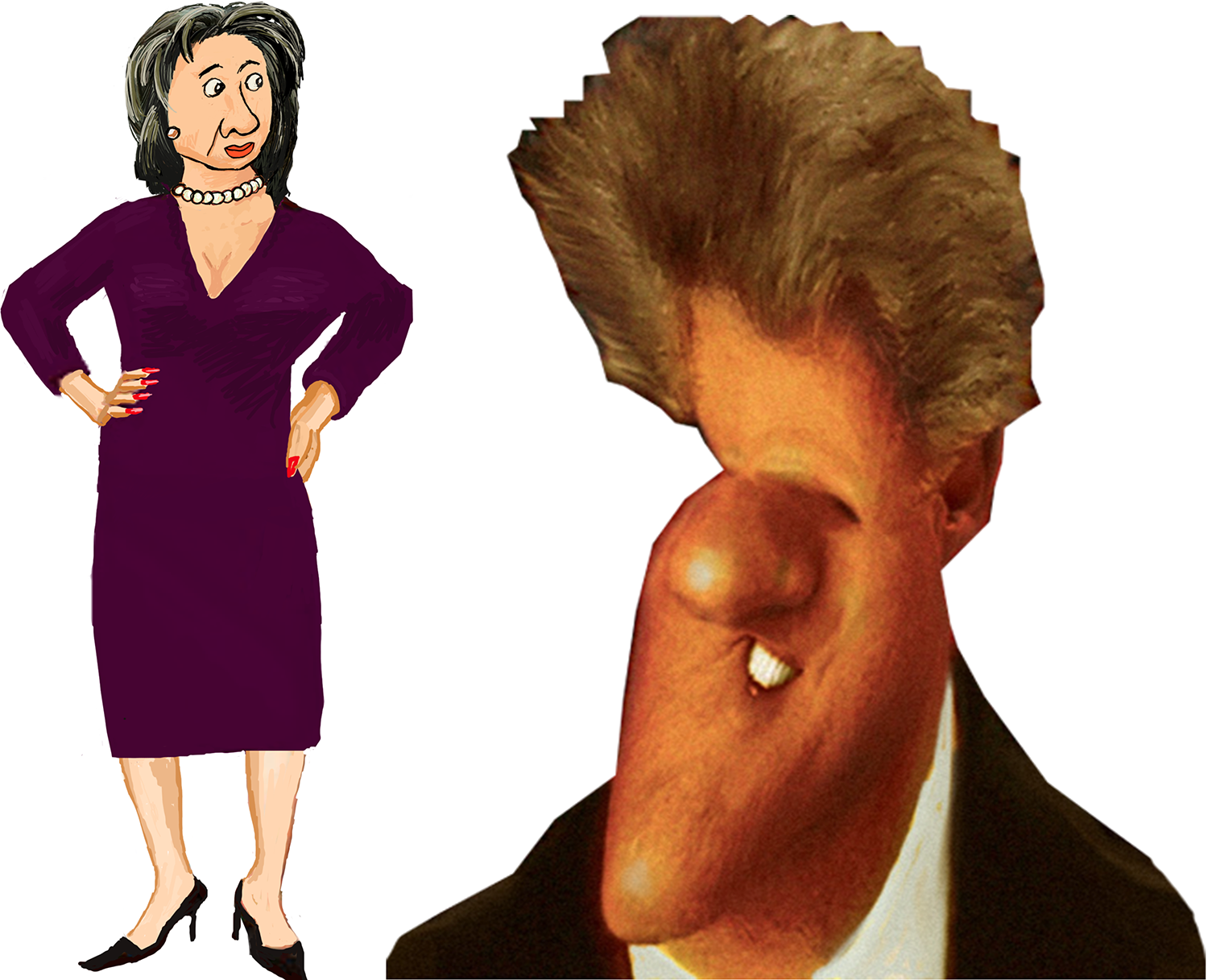}}&
       \hfill
		\fbox{\includegraphics[width=0.53\linewidth]{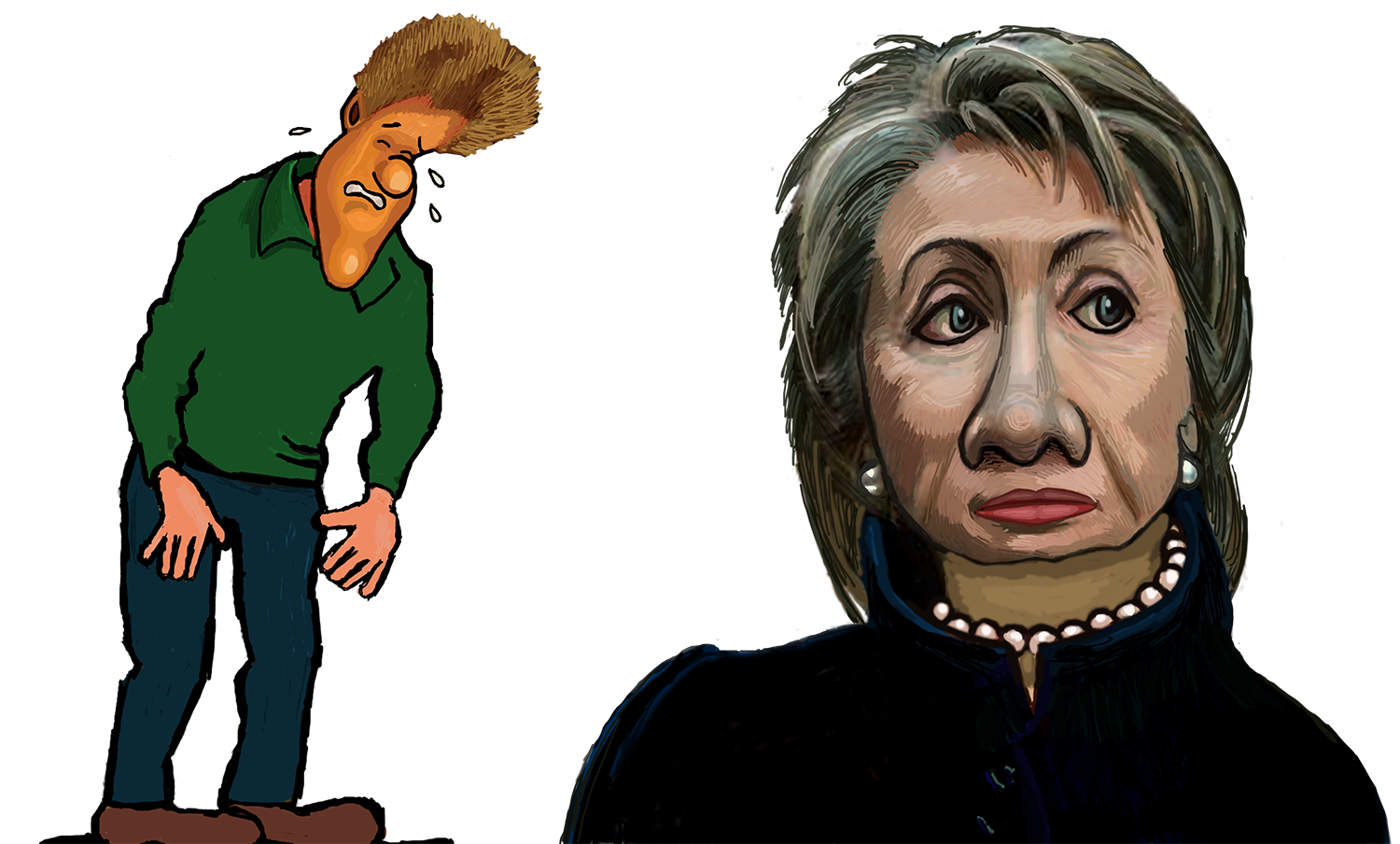}}\\
    \end{tabular}
		\caption{\it Two examples that demonstrate objects cut from the bottom can be used as visually valid compositions if they are placed at the bottom. These frames are considered Apex, although they are really external. }
		\label{fig:Acceptable2}	
	\begin{tabular}{cccc}
		\fbox{\includegraphics[width=0.44\linewidth]{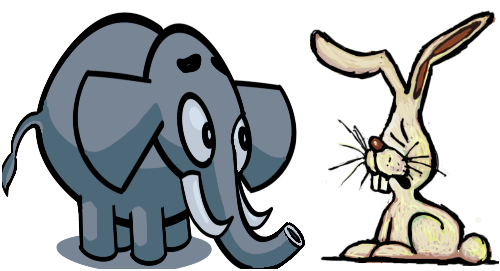}}&
		\hfill
       \fbox{ \includegraphics[width=0.50\linewidth]{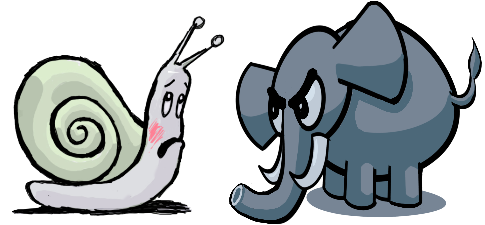}}\\
    \end{tabular}
		\caption{\it Two examples that demonstrate the actual objects' sizes do not matter to obtain visually valid compositions if we place the bottoms of both images approximately at the same level. These views are actually external. However, we consider them Apex. }
		\label{fig:Acceptable4}	
			\begin{tabular}{cccc}
       \fbox{ \includegraphics[width=0.45\linewidth]{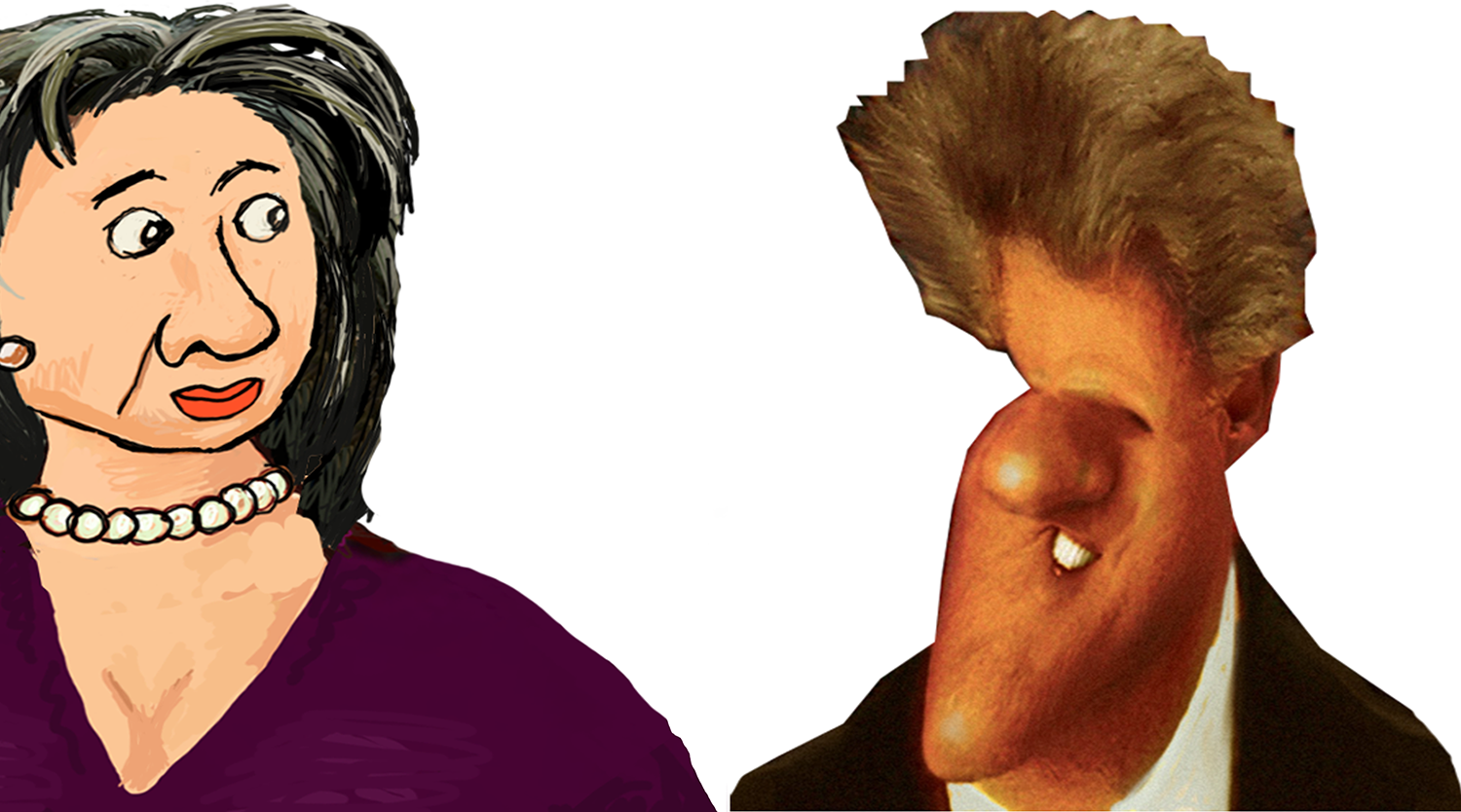}}&
       \hfill
		\fbox{\includegraphics[width=0.45\linewidth]{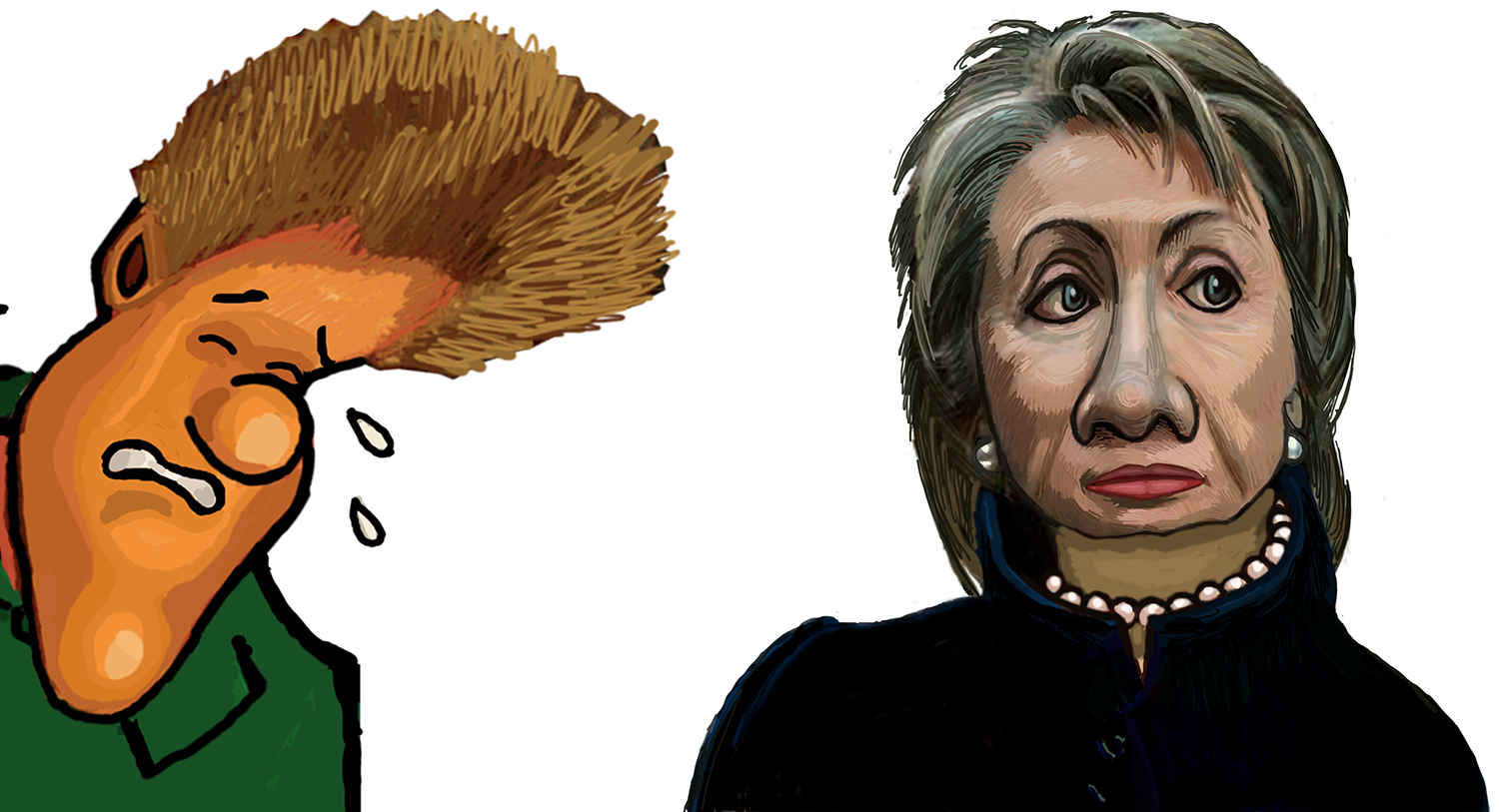}}\\
    \end{tabular}
		\caption{\it Actual Apex views corresponding to characters in Figure~\ref{fig:Acceptable2}. We do not consider this Apex since the bottoms of images are not at the same level. Note that the images at the left are cropped. Their bottom positions are below. }
		\label{fig:Acceptable3}	
\end{figure}

Our key observation comes from a fundamental perspective rule. If there is a ground plane and all objects/characters stand on this plane, the bottom of the objects that are closer to the camera will be lower in the image. Therefore, the placement in  Figure~\ref{fig:Unacceptable} is not acceptable since the characters that appear closer to the camera stay higher in the image. On the other hand, the bottom positions can be the same regardless of their closeness to the camera. We can always obtain this acceptable solution by putting the camera exactly in the bottom plane as shown in Figure~\ref{fig:Acceptable}. This method also works with objects cut from the bottom as shown in Figure~\ref{fig:Acceptable2}. Note that the placements shown in Figures~\ref{fig:Acceptable} and~\ref{fig:Acceptable2} are actually external views since one of the objects/characters are closer to the camera than the other. Moreover, the sizes of objects/characters also matter to classify a view as Apex as shown in Figure~\ref{fig:Acceptable4}. On the other hand, since we do not have any information about physical objects/characters that images represent, it is not possible to identify exact Apex views. Therefore, our solution is to classify all cases in which the bottoms of the two images are exactly in the same position as the Apex view. 	

Our method is not really an interpolation of two end views, but it is an extrapolation of an Apex view into both external directions. 
We start with an acceptable Apex position. We then extrapolate this Apex view towards both the external right and external left. This extrapolation can simply be done by continuous translation and scaling. In the next section, we present a special Apex view without a loss of generality. 

\begin{figure}[htbp!]
	\begin{tabular}{cccccccccc}
		\includegraphics[width=0.45\linewidth]{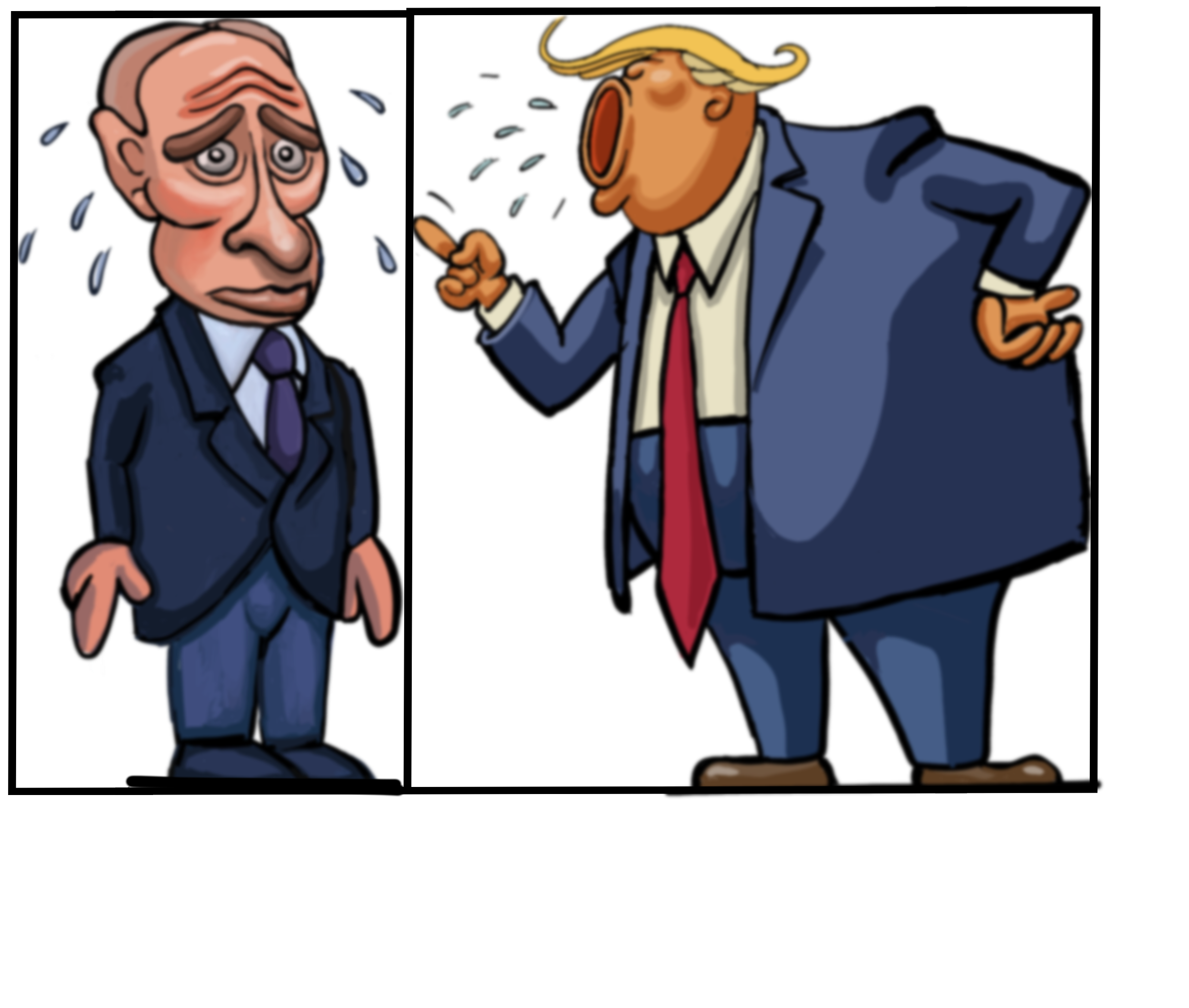}&
		\hfill
       \includegraphics[width=0.45\linewidth]{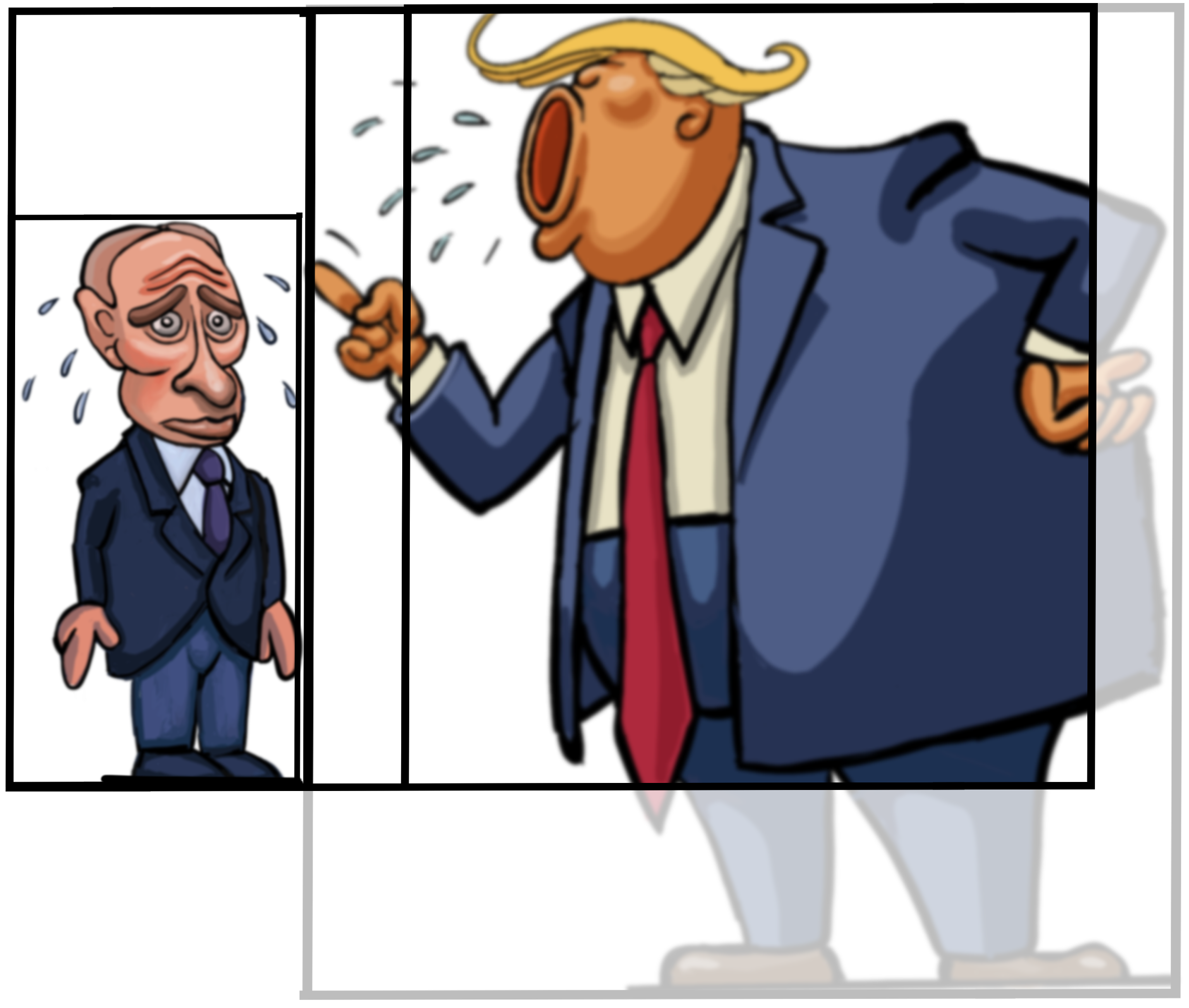}\\
         (A) Apex View. &
     (B) External right.  \\
    \end{tabular}
		\caption{\it The specific Apex view that we use in the presentation of the methodology and an Exterior view that is obtained by scaling, translating and cropping the two images. }
		\label{fig:Apex}	
\end{figure}

\section{Methodology \label{sec:methodology}}

We present our extrapolation process starting with the Apex view into two images that have the same height and that are placed side by side without a gap as shown in Figure~\ref{fig:Apex}(A). We can always make any two images the same height, say $H$ by resizing them appropriately. Let these two images be denoted by $I_{1}$ and $I_{2}$; and their initial width and height be denoted by $(w_{1},h_{1})$ and $(w_{2},h_{2})$ respectively. Now, let $s_1$ and $s_2$ denote uniform scaling parameters with the property $s_1 = H/h_{1}$ and  $s_2=H/ h_{2}$, we can form a new composite image with width and height and $W=s_1 w_{1} + s_2 w_{2}$ and $H=s_1 h_{1} = s_2 h_{2}$. This gives us a well-defined Apex view panel shown at the left in Figure~\ref{fig:Apex}(A). 

Now, to produce an external view, we scale down one of the images, say $I_{1}$. Let $t_1$ denote the new scaling parameter, then the new width of image 1, $w'_1$ becomes $w'_1= t_1 s_1 w_1$. This leaves a space between image 1 and image 2. We can scale up image 2 just to fill the space. However, the external view requires that a part of the image goes out of the panel. If we scale to fill the gap as shown in Figure~\ref{fig:Apex}(B), we obtain an exterior view. The gap created by scaling down image 1 is $s_1 w_1 - w'_1 = (1-t_1) s_1 w_1$. As a result, the width of image 2 must be at least $s_2 w_2 + (1-t_1) s_1 w_1$. In other words, the scaling $t_2$ must be at least the following:
$$t_2 \geq \frac{s_2w_2 + (1-t_1) s_1 w_1}{s_2 w_2}= 1 + \frac{(1-t_1)s_1 w_1}{s_2 w_2}$$
In this case, to make image 2 cropped on the right side, we use the following equation for scaling image 2:
$$t_2 = 1 + \frac{a(1-t_1)s_1 w_1}{s_2 w_2}$$
where $a$ is a number larger than 1. In our implementation, we use $a=2$ to make the images cropped exactly the same amount as the new space created as shown in Figure~\ref{fig:Apex}(B).

We never crop the top of the image, although cropping the top is common in movies and comics. To keep the top of the image at exactly the same level, we simply scale image 2 using the upper left corner as the origin of scaling. To remove the gap between the two images, we further translate image 2 in the $x$ direction, in the following amount: 
$$x_2 = - (1-t_1) x_1 w_1$$
An important property of this operation is that it is independent of what we choose for $a$ and $t_2$. So, a single parameter $t_1$ between $0$ and $1$ is sufficient to produce extrapolation from the Apex view to the External Right view. Internal Right is obtained in the limit when $t_1=0$. 

For extrapolation from Apex to External Left view, we change $t_2$ from $1$ to $0$. We compute the scaling and translation of image 1 in a reciprocal way. We compute $t_1$ using $t_2$ and $a$ as  
$$t_1 = 1 + \frac{a(1-t_2)s_2 w_2}{s_1 w_1}$$
Using $t_1$, we scale image 1 upper right corner of image 1 as the origin of the scaling. Then, to remove the gap between the two images, we translate the image 1 in the $x$ direction by the following amount: 
$$x_1 = - (1-t_2) x_2 w_2$$
In this case, a single parameter $t_2$ between $0$ and $1$ is sufficient to produce an extrapolation from the Apex view to the External Left view. Internal Left is obtained in the limit when $t_2=0$.  

This idea can be extended to other Apex views. Note that it is also possible to create a wide variety of width/height ratios for Apex views by choosing $s_0 = H/h_{0}$,  $s_1 < H/ h_{1}$ and given a $w_0$ amount of gap between two images. Then $W=w_0+ s_1 w_{1} + s_2 w_{2}$. The previous equations will still hold for these cases with some minor adjustments. This flexibility is particularly useful to produce a comic page that consists of multiple panels. This is why it is easy to design comic pages.

Multi-object/character external views can simply be obtained by scaling and translating only the characters in the extreme right and extreme left in the original Apex views. 

\section{Implementation}

We have implemented this single parameter interface as a single-panel cartoon creation tool on the Web using JavaScript and webGL. This tool was developed to provide only a minimal amount of control to users. We have only one slider to obtain all views from Apex to External and Internal. The advantage of minimal control is that the user need not worry about too many settings in order to obtain the desired aesthetic. 

We have provided multiple expressions for objects/characters such that the users can choose different emotions that can be appropriate for the text. We have also included sentiment analysis to choose appropriate expressions for a given text. However, sentiment analysis currently works only offline so has not yet been included in the Web-based system. We allow horizontal flips to make two characters look at each other or different directions. Such flips can also change the overall impression of dynamics between two objects/characters. We have also added word balloons that point to the images. 

Using this tool as an underlying infrastructure, we have provided four different selections for creating science, mathematics, animal, and political cartoons. Each uses a different database of images. We allow to use of all images together. Our web-based system for single-panel cartoons can be viewed at http://storytelling.viz.tamu.edu. 

\begin{figure*}[htbp!]
		\fbox{\includegraphics[width=0.95\linewidth]{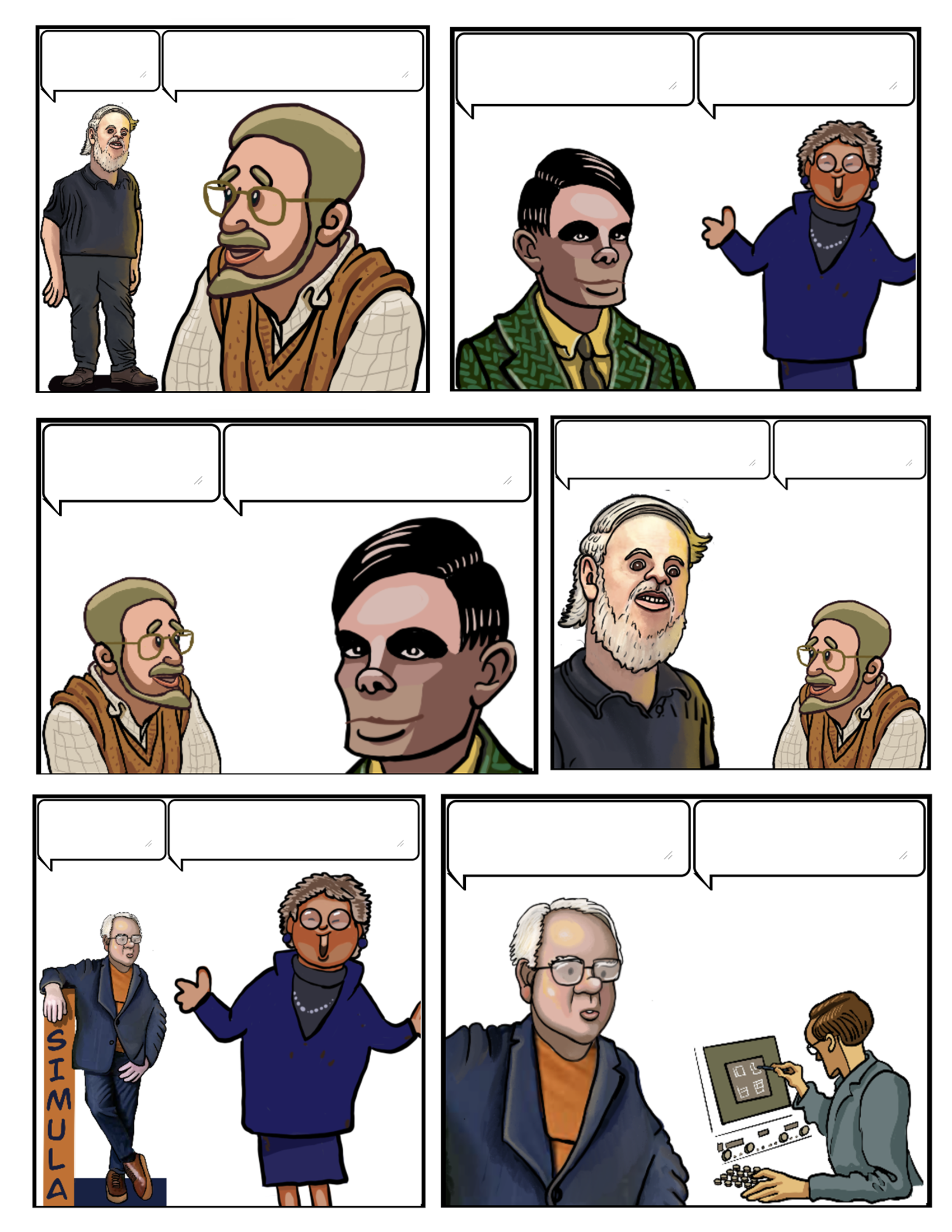}}
		\caption{\it A potential comic page featuring Frances Allen, Edsger W. Dijkstra, Dennis Ritchie, Ivan Sutherland, and Alan Turing. We have composited the $8.5\times11$ page using single cartoon panels. One can try to fill out balloons to see if it can turn into a meaningful story.    }
		\label{fig:page1}	
\end{figure*}

Having single panels is useful for creating comic pages. Figures~\ref{fig:page1},~\ref{fig:page2} and~\ref{fig:page3} show three comic pages constructed from single-panel cartoons generated by our prototype. A comic artist randomly created the panels in the prototype system and then put the individual panels together in order in Photoshop resizing as needed to make a row's pair of panels the same height and the rows the same width but without changing the width and height ratios of the generated panels. This exercise suggests that such pages can eventually be created automatically by packing rectangles appropriately. Although the general problem of 2D packing problems is hard \cite{lodi2002two}, this particular case has extra flexibilities and constraints that simplify the problem. It is even possible to easily change the width and height ratios of panels by changing the distances between the characters and the number of characters. Our next goal is to provide full pages automatically. 

Our experience is that once a page is provided almost anybody can come up with an interesting story just by filling out balloons. We conjecture that the inherent knowledge about the characters plays an important role in the creation of these stories. The reader is invited to try to fill in the balloons in Figures~\ref{fig:page1},~\ref{fig:page2} and~\ref{fig:page3} to test this conjecture. In the prototype, we also provided links for each character to their corresponding Wikipedia page to enable viewers to obtain more information. 
Another feature of this system is that information on each artwork can be represented in a simple database. This information, in particular, can include the name of the creator. In the current system, all images are created by only one professional cartoonist. We think that is important to motivate other artists to contribute with their work since they will be recognized.

\section{Discussion and Future Work}

In this work, we have demonstrated that one continuous parameter can be used to obtain visually valid placement of characters in single cartoon panels. This parameter can be considered a continuous interpolation from five qualitatively different 3D cameras for two characters: Internal Left, External Left, Apex, External Right, and Internal Right views \cite{he1996virtual}. These five views are widely used in movies and comics to create dynamic dialogues \cite{arijon1991grammar}. We have also demonstrated that we can obtain a significant range of width and height ratios for composited panels. These results suggest that the actual design space is significantly rich and it is possible to easily obtain visually valid cartoon frames. 

It is important to note that we do not create all visually valid compositions. We only create some compositions that are guaranteed to be visually valid. It is possible to provide more degrees of freedom. For instance, we can allow overlapping the characters with an additional slider. Since we know which character is front, we can order them in space. However, can cause some non-valid compositions. It is hard to guarantee to obtain only valid cases. 

This type of limitation also explains the difficulty faced by naive users. Potential placements of objects/characters in a given cartoon frame define a very large design space. Most of these placements are useless. If you randomly search for a solution without the experience, it is impossible to find a valid one. Therefore, cartoon expertise is critical to obtain visually valid solutions. We have developed this solution with significant input from a professional cartoonist. 

\begin{figure*}[htbp!]
		\fbox{\includegraphics[width=0.95\linewidth]{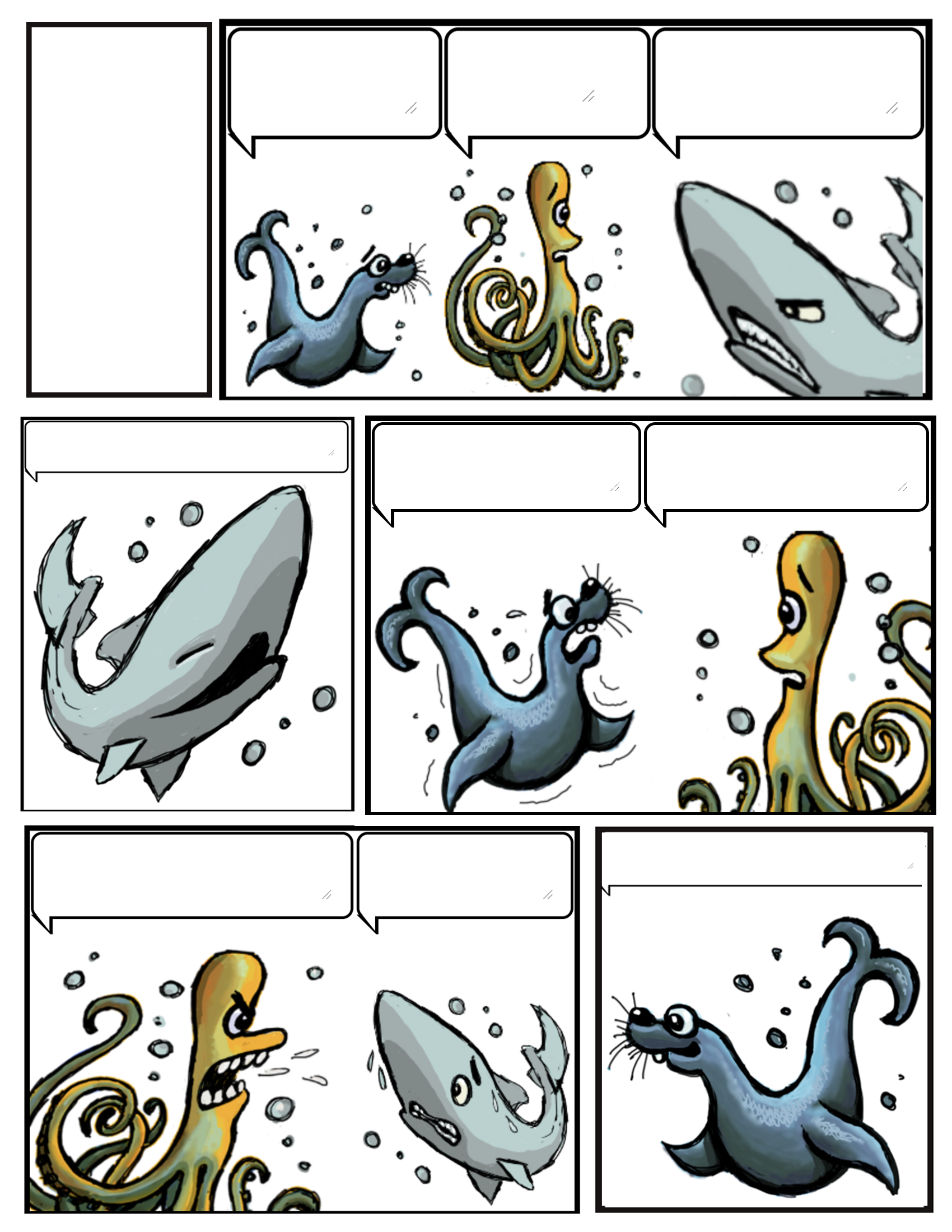}}
		\caption{\it Another potential comic page featuring an octopus, a sea lion, and a shark. We again have composited the $8.5\times11$ page using single cartoon panels. Note that we added an empty rectangle for the ``title'' page, which also simplifies the rectangle packing problem.    }
		\label{fig:page2}	
\end{figure*}

\begin{figure*}[htbp!]
		\fbox{\includegraphics[width=0.95\linewidth]{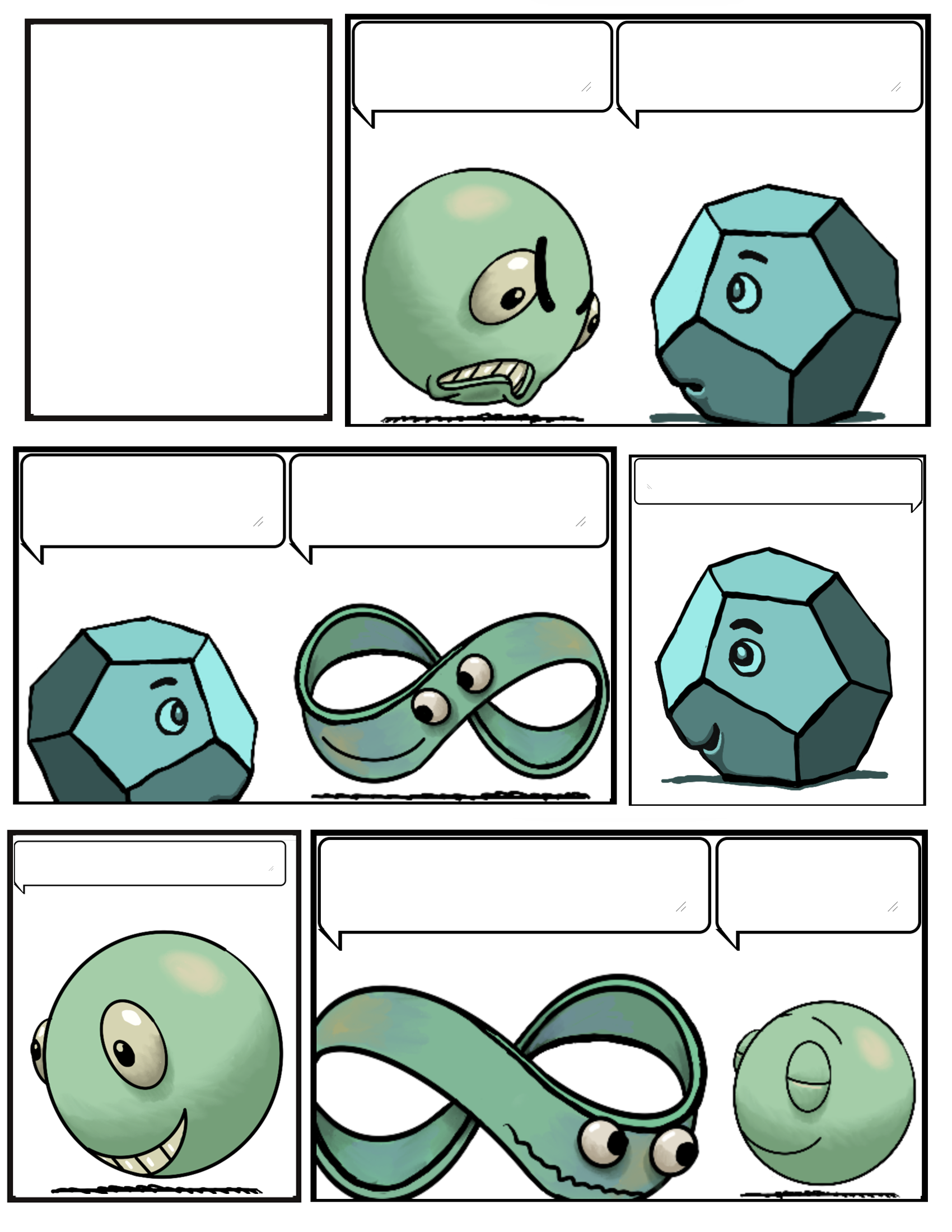}}
		\caption{\it Another potential comic page featuring a dodecahedron, a sphere, and a Moebius strip. We again have composited the $8.5\times11$ page using single cartoon panels. Note that we added an empty rectangle for the ``title'' page, which also simplifies the rectangle packing problem.    }
		\label{fig:page3}	
\end{figure*}

Figures~\ref{fig:page1}~\ref{fig:page2}  and~\ref{fig:page3} are good examples of the importance of cartoon and comic experience. It did not take much time for a cartoonist to create those pages from a set of 6-7 cartoon panels. There was only a need to decide which panels would come next to each other. Then, it was possible to quickly put together an $8.5\times11$ comic page. Here the main constraint is the size of the page. The panels are chosen randomly just to fit this space. Despite that, it is clear that some sort of story emerges. 

This particular solution for single-panel cartoon creation is developed to serve as an infrastructure for the automatic creation of comic pages that consist of multiple panels. Our results demonstrate that it is possible to randomly create single panels that can fit nicely into any given comic page format from $8.5\times11$ to A4. We can also use a Markov model to pick up expressions automatically \cite{liu2012never}. This work also suggests that it is possible to create panels with changing camera views that can provide a dynamic look. Such dynamism is critical for effective storytelling. It can also be possible to create these panels automatically from a given text. 

We can also choose expressions automatically using sentiment analysis. We initially added sentiment analysis to obtain character expressions automatically based on text. However, we could not include it web-based interactive tool as our initial attempts made the interface a little too complicated. On the other hand, we can later include sentiment analysis.

In this system, all characters and their expressions are created only by one professional cartoonist. Our current implementation also allows the inclusion of other artists by providing acknowledgment per image. We think that we have to rely on artists since It is still hard to automatically create very expressive characters. On the other hand, we can still use automatically created characters even if they are not that expressive. This will be helpful to turn texts into comic stories automatically. Such a system can also be used to create storyboards from screenplays for movies. Such storyboards can be used to identify weaknesses in stories. 

One of the main problems is the creation of expressive balloons. Comic book artists spend a significant amount of time placing balloons. The placement and orientation of balloons play an important role in easy-to-read panels. Our current balloon placement solution is visually valid, but it is very conservative and not visually interesting.

\bibliographystyle{unsrtnat}
\bibliography{references}

\end{document}